\documentclass[12pt]{article}
\usepackage{cite}
\usepackage[T1]{fontenc}
\usepackage[utf8]{inputenc}

\usepackage[english]{babel}

\usepackage{graphicx}
\usepackage[dvipsnames]{xcolor}
\usepackage{multirow}
\usepackage[thinlines]{easytable}
\usepackage{amsmath}
\usepackage{amsfonts}
\usepackage{amssymb}
\usepackage{amstext}
\usepackage{cancel}
\renewcommand{\text}[1]{%
\ifthenelse{\equal{#1}{mB}}{m_B}{}%
\ifthenelse{\equal{#1}{mc}}{m_c}{}%
\ifthenelse{\equal{#1}{sbar}}{\bar{\sigma}}{}%
\ifthenelse{\equal{#1}{fB}}{f_B}{}%
\ifthenelse{\equal{#1}{q2}}{q^2}{}%
}
 
\allowdisplaybreaks
\usepackage{slashed}
\usepackage{braket}
\usepackage{tablefootnote}
\usepackage[bookmarks, breaklinks, colorlinks,urlcolor=black, citecolor=red, 
linkcolor=blue]{hyperref}

\numberwithin{equation}{section}

\usepackage{multirow}
\usepackage{booktabs}

\usepackage{hyperref}
\usepackage{placeins}
\usepackage{verbatim}

\usepackage[normalem]{ulem}
\usepackage{cancel}

\usepackage{float} 

\def\bc{\begin{center}}
\def\ec{\end{center}}
\def\be{\begin{equation}}
\def\ee{\end{equation}}
\def\bea{\begin{eqnarray}}
\def\eea{\end{eqnarray}}

\def\gev{\ensuremath{\mathrm{Ge\kern -0.1em V}}}
\def\mev{\ensuremath{\mathrm{Me\kern -0.1em V}}}

\newcommand{\OPE}{\textrm{OPE}}
\newcommand{\HAD}{\textrm{had}}

\newcommand{\fit}{\textrm{fit}}
\newcommand{\LO}{\textrm{LO}}
\newcommand{\NLO}{\textrm{NLO}}
\newcommand{\pert}{\textrm{pert}}
\newcommand{\cond}{\textrm{cond}}
\newcommand{\cont}{\textrm{cont}}
\renewcommand{\th}{\textrm{th}}

\newcommand{\MeV}{\textrm{\,MeV}}
\newcommand{\GeV}{\textrm{\,GeV}}
\newcommand{\B}{\mathcal{B}}
\newcommand{\C}{\mathcal{C}}
\newcommand{\D}{\mathcal{D}}
\newcommand{\E}{\mathcal{E}}
\newcommand{\F}{\mathcal{F}}
\newcommand{\T}{\mathcal{T}}

\newcommand{\Fi}{\F^{(i)}}
\newcommand{\Fp}{\F^{(p)}}
\newcommand{\Fq}{\F^{(q)}}

\newcommand{\rhoi}{\hat{\rho}_{\cont}^{(i)}}
\newcommand{\rhop}{\hat{\rho}_{\cont}^{(p)}}
\newcommand{\rhoq}{\hat{\rho}_{\cont}^{(q)}}
\newcommand{\rhomu}{\hat{\rho}_{\cont,\mu}}

\definecolor{darkgreen}{RGB}{30,150,30}

\newcounter{TODO}

\newcommand{\refapp}[1]{Appendix~\ref{app:#1}}
\newcommand{\reffig}[1]{Fig.~\ref{fig:#1}}
\newcommand{\refeq}[1]{Eq.~(\ref{eq:#1})}
\newcommand{\refeqs}[2]{Eqs.~(\ref{eq:#1})-(\ref{eq:#2})}
\newcommand{\refsec}[1]{Section~\ref{sec:#1}}
\newcommand{\refsubsec}[1]{Subsection~\ref{sec:#1}}
\newcommand{\reftab}[1]{Table~\ref{tab:#1}}

\begin{document}

\begin{flushright}
SI-HEP-2023-21\\
P3H-23-064 
\end{flushright}
	
\renewcommand*{\thefootnote}{\fnsymbol{footnote}}

\vskip 2cm	
	
	\begin{center}
		
		{\Large\bf \boldmath $B\to D_0^*$ and $B_s\to D_{s0}^*$ form factors \\[2mm] from QCD light-cone sum rules} \\[6mm]
	{
	Nico
	Gubernari$^a$\footnote{Email: nico.gubernari@gmail.com}, 
	Alexander Khodjamirian$^a$\footnote{Email: khodjamirian@physik.uni-siegen.de}, 
    Rusa Mandal$^b$\footnote{Email: rusa.mandal@iitgn.ac.in},
		and  Thomas Mannel$^a$\footnote{Email: mannel@physik.uni-siegen.de }
}
\\[6pt]
	
	$^a${\small\it Center for Particle Physics Siegen (CPPS), Theoretische Physik 1,
           \\ Universit\"at Siegen, 57068 Siegen, Germany}

   $^b${\small\em Indian Institute of Technology Gandhinagar, Department of Physics, \\ Gujarat 382355, India}

	\end{center}
	
	%

\begin{abstract}
\noindent
We present the first application of QCD light-cone sum rules (LCSRs) with  $B_{(s)}$-meson distribution amplitudes to the $B_{(s)}\!\to\! D_{(s)0}^*$ form factors, where
$D_{(s)0}^*$ is a charmed scalar meson. 
We consider two scenarios  for the $D_0^*$  spectrum.
In the first one, we follow the Particle Data Group and consider a single broad resonance $D_0^*(2300)$. 
In the second one, we assume the existence of two  scalar resonances, $D_0^*(2105)$ and $D_0^*(2451)$, as follows  from a recent theoretically motivated analysis of $B\to D\pi\pi$ decays.
The $B\!\to\! D_0^*$  form factors are calculated in both scenarios, also taking into account the large total width  of $D_0^*(2300)$. 
Furthermore, we calculate the $B_s\!\to\! D_{s0}^*$  form factors, considering in this case only the one-resonance scenario  with $D_{s0}(2317)$.
In this LCSRs calculation, the $c$-quark mass
is kept finite and the $s$-quark mass is taken into account.
We also include contributions of the two- and three-particle distribution amplitudes up to twist-four. 
Our predictions for semileptonic $B\!\to\! D_0^*\ell\nu_\ell$ and $B_s\!\to\! D_{s0}^*\ell\nu_\ell$ branching ratios are compared with the available data and HQET-based
predictions.
As a byproduct, we also obtain the $D_0^*$- and $D_{s0}^*$-meson decay constants
and predict the lepton flavour universality ratios $R(D_0^*)$ and $R(D_{s0}^*)$.

\end{abstract}

\renewcommand*{\thefootnote}{\arabic{footnote}}
\setcounter{footnote}{0}

\newpage


{\hypersetup{linkcolor=black}\tableofcontents}

\section{Introduction}
\label{sec:intro}

The semileptonic $b\to c \ell \bar{\nu}_\ell$ transitions
are predominantly realised in the form of exclusive 
$\bar B\to D\ell\bar{\nu}_\ell$ and $\bar B\to D^*\ell\bar{\nu}_\ell$ decays and their $\bar{B}_s$ counterparts. 
No less important but considerably less studied are the subdominant exclusive decays $\bar{B}\to D^*_{J}\ell\bar{\nu}_\ell$  and $\bar{B}_s\to D^*_{sJ}\ell\bar{\nu}_\ell$, where  $D_J^*$ and $D_{sJ}^*$ are excited charmed mesons with spin-parity $J^P=0^+,1^+,2^+$.
Apart from filling the gap between  the inclusive $\bar{B}\to X_c\ell\bar{\nu}_\ell$ width and the sum over partial widths of exclusive channels,  
these decays are important also for deciphering the spectroscopy of excited charmed
mesons which is still far from being firmly established. 
Indeed, currently, the main information
on the masses and widths of 
$D_J^*$ mesons stems from the analyses of the nonleptonic three-body $B$-decays in which one has to disentangle complicated final-state interactions.
Semileptonic decays such as $\bar{B}\to D^{(*)}\pi \ell\bar{\nu}_\ell$  are in this respect much simpler for an extraction of $D_J^*$ resonances in the $D^{(*)}\pi$ states. 
It is also possible to use the $\bar{B}\to D^*_{J}\ell\bar{\nu}_\ell$ decays as an alternative channel to probe the $b\to c \ell \bar{\nu}_\ell$ transitions for the presence of  New Physics. 
To this end, however, one would need more precise experimental measurements and theoretical predictions. 
In addition, these decay channels are important because they constitute one of the main backgrounds in the $\bar{B}\to D^*\ell\bar{\nu}_\ell$ measurements.

The key element needed to obtain predictions for any exclusive semileptonic decay is a set
of process-dependent hadronic form factors. 
The lattice QCD methods, well advanced in calculating the $B\to D$ and $B\to D^*$ form factors, are not yet able to describe the $B$-meson transitions to unstable charmed mesons, 
especially to the broad resonant $D^{(*)}\pi$ states with $J^P=0^+,1^+$. 
In our previous paper \cite{Gubernari:2022hrq}, we obtained the form factors of the $B$-meson transitions to the charmed axial mesons with $J^P=1^+$, applying the QCD light-cone sum rules (LCSRs) with $B$-meson distribution amplitudes (DAs). 
This version of the LCSR method is very versatile, as it allows the spin parity and flavour of the final hadronic state to be varied by adjusting the interpolating quark-antiquark current in the underlying correlation function. 

In this paper we use LCSRs with $B$-meson DAs
to calculate for the first time 
the $B\to D_0^*$ and $B_s\to D_{s0}^*$ form factors, including the contributions of three-particle $B$-meson DAs up to twist-four.\footnote{ 
    Earlier, the three-point QCD sum rules  based on the local OPE and  double dispersion relation have been used for the $B\to D_0^*$ form factors in Ref.~\cite{Colangelo:1991ug} (see also the HQET analogue of these sum rules in Ref.~\cite{Colangelo:1998ga}).  
}
Important additional elements of this
paper are also the two-point QCD sum rules 
for the decay constants of $D_0^*$. 
The correlation functions
used here and in Ref.~\cite{Gubernari:2022hrq} are  
very similar, up to a replacement of the charmed-meson interpolating current.  

In the case of charmed axial  mesons, there was a  problem  
to separate the form factors for two very close $D_1^*$ resonances. The solution found  in Ref.~\cite{Gubernari:2022hrq} 
was to introduce a second interpolating current
and to use linear combinations of LCSRs and of two-point sum rules with  two different currents. 
In the case of charmed scalar mesons the hadronic parts 
of the sum rules do not pose such a problem. 
Instead, we are faced with two alternatives concerning the experimentally observed charmed scalar mesons. 
According to Ref.~\cite{ParticleDataGroup:2022pth},
the lightest meson is identified with the broad $D_0^*(2300)$ resonance decaying into $D\pi$ in the $S$ wave. 
On the other hand,  in recent theory-based analysis of $B\to D \pi\pi$ decays (see Refs.~\cite{Du:2020pui,Du:2017zvv}
and references therein) 
a different configuration of charmed scalar mesons 
was found, consisting of the two well separated  resonances, $D_0^*(2105)$ and  $D_0^*(2451)$. 
In particular, the 
lowest state $D_0^*(2105)$ is  interpreted as a product of nonperturbative dynamics of 
the low-energy pion scattering off $D$-meson, analogous to the 
lightest  scalar nonstrange and strange mesons, $f_0(600)$ and $K^*_0(700)$. The 
latter are usually interpreted (see, e.g. the review on scalar mesons in Ref.~\cite{ParticleDataGroup:2022pth})
as  molecular and/or tetraquark objects rather than quark-antiquark states. 
The existence of a lighter charmed scalar meson than the one identified in 
Ref.~\cite{ParticleDataGroup:2022pth} was also supported by a recent lattice QCD computation \cite{Gayer:2021xzv}
of the isospin 1/2  $D\pi$ scattering amplitudes.

We  will consider both the one-resonance and the two-resonance scenarios for the charmed scalar mesons and compute the respective decay constants and form factors. 
For the charmed-strange scalar meson we limit ourselves to the single resonance scenario, since the ground state is narrow and well established experimentally.

The plan of this paper is as follows.
In Section~\ref{sec:reson}, we introduce and discuss both scenarios of the spectrum of   charmed and charmed-strange scalar mesons. In Section~\ref{sec:2ptSR}, we use the two-point QCD sum rules to obtain the
decay constants of these mesons. Section 
\ref{sec:lcsr} is devoted to the LCSRs
for the form factors and their numerical analysis, including the prediction of selected physical observables. 
Finally, section \ref{sec:concl} contains the concluding discussion.
The paper has two appendices: in \ref{app:DA} we collect 
the definitions and the models for the $B$-meson light-cone DAs  
and in \ref{app:OPE} we present the expressions for the OPE coefficients of our LCSRs.

\section{Charmed scalar mesons}
\label{sec:reson}

For the QCD sum rules and LCSRs to be obtained below  we need as an input the 
masses and the widths of the lowest charmed and  charmed-strange scalar mesons. 
Here we discuss the two alternatives already mentioned in the Introduction.
As a  default choice (scenario 1), we adopt  the single resonance $D^*_0(2300)$ 
as it is currently listed in Ref.~\cite{ParticleDataGroup:2022pth}. 
A rather strong argument against this choice is 
that in this case the corresponding charmed-strange meson $D^*_{s0}(2317)$ 
seems to be unnaturally light. 
Indeed,  the 
mass difference between strange and nonstrange resonances 
does not fit the expected order of  $m_s$.
Note that,  according to Ref.~\cite{ParticleDataGroup:2022pth},
the mass and width of the
$D^*_0(2300)$ resonance  is obtained from Dalitz-plot 
analysis
of the weak  $B\to D\pi\pi$ decays. 
The mass of its strange counterpart $D^*_{s0}(2317)$ is, on the contrary,  
more directly determined  from  the $e^+e^-\to D_{s0}^*\bar{D}_s$ 
cross section with the subsequent $D^*_{s0}\to D_s \pi^0$ decay
 \cite{BESIII:2017vdm} .
 
The second alternative for charmed scalar  mesons 
is markedly different and originates from the analysis done in Refs.\cite{Du:2020pui,Du:2017zvv},
and in  earlier papers cited therein. 
Here, again the data on  $B\to D\pi\pi$ decay are used, more specifically,
the most accurate recent measurements by LHCb \cite{LHCb:2016lxy}. 
Without going into details 
which are beyond the scope of this paper, we only mention that in this analysis 
the  $S$-wave $D\pi$ scattering amplitude is isolated 
in the final-state interaction of the $B\to D\pi\pi$ decay and the resonance structure
of this amplitude is extracted. 
The outcome is a prediction of two resonances 
$D^*_0(2105)$ and $D^*_0(2451)$ replacing the single one suggested in 
Ref.~\cite{ParticleDataGroup:2022pth}. 
The first of these resonances solves the above mentioned problem of the mass difference between strange and nonstrange states.
Simultaneously, the second one indicates the existence of a second charmed-strange state, the analog of  $D^*_0(2451)$. Adding to its mass the  difference between the
$D^*_{s0}(2317)$ and $D^*_0(2105)$ masses, we roughly locate this state at around 2660 MeV.

It is very important to reestablish 
or test these predictions in a more clean hadronic environment 
provided by semileptonic decay $B\to D\pi \ell \nu$. To this end,
one needs an accurate partial wave reconstruction of the $D\pi$ state,  
an isolation of the S-wave component and a study of its resonance
structure. Needless to say, a theory prediction for the underlying 
$B\to D_0^*$ transition form factors is important for such an analysis.

\begin{table}[t]
\def\arraystretch{1.5}
	\begin{center}
		\begin{tabular}{|c|c|c|c|}
			\hline 
			Scenario & Meson & Mass [\mev] & Width [\mev] \\ 
			\hline\hline \noalign{\vskip2pt}
			\multirow{2}{*}{$1$} 
			& $D_0^* \equiv D_0^*(2300)$ & $ 2343\pm 10$    & $229\pm 16$ \\
			& $D_{s0}^*\equiv D_{s0}^*(2317)$ & $2317.8\pm 0.5$ & $ <3.8$ \\
			\hline
			\multirow{4}{*}{$2$} 
			& $D_0^* \equiv D_0^*(2105)$  & $2105_{+6}^{-8}$    & 
   $204_{-22}^{+20}$   \\
			& $D_0^{*\prime} \equiv D_0^*(2451)$   & $2451_{+35}^{-26}$    & 
      $ 268_{-16}^{+14}$     \\
			& $ D_{s0}^*\equiv D_{s0}(2317)$  & $ 2317.8\pm 0.5$ & $<3.8$ \\
			& $D_{s0}^{*\prime}\equiv D_{s0}^*(2660)$ & $\sim 2660$ & --- \\
			\hline
		\end{tabular}
		\caption{
		\emph{The lowest-lying charmed scalar ($J^P=0^+$) mesons. 
            For scenario 1 (scenario 2) we take the masses and total widths from Ref.~\cite{ParticleDataGroup:2022pth}  (from Refs.~\cite{Du:2017zvv,Du:2020pui}), except the mass of $D_{s0}^{*\prime}$ which is our estimate. 
	}}
  \label{tab:spectr}
\end{center}
\end{table}

\section{Two-point QCD sum rules for charmed scalar mesons}
\label{sec:2ptSR}

The decay constants of the charmed scalar  meson and its charmed-strange counterpart are essential inputs for the LCSRs for the $B_{(s)}\!\to\! D_{(s)0}^*$ form factors.
Since there is no available estimate of these quantities and in particular no lattice QCD computation, we calculate them using QCD sum rules.
We derive these sum rules starting from the two-point correlators
\begin{eqnarray} 
    \label{eq:corr2pt}
    \Pi^{(s)}  (q^2) = i \int d^4 x \, e^{iqx} \langle 0 | \mathcal{T}\{ J_{(s)} (x) J_{(s)}^{\dagger}(0)   \} | 0 \rangle 
    \,,
\end{eqnarray} 
where
\begin{align}
\label{eq:J}
    J = (m_c - m_d) \bar{c} d ~~~
\mbox{and} ~~~    J_s=(m_c-m_s)\bar{c}s 
\end{align}
are the interpolating quark currents for the $D_0^*$ and $D_{s0}^*$ meson, respectively. 
The currents in \refeq{J} coincide with the divergences of the corresponding vector currents.
These currents have no anomalous dimension. 
We assume isospin symmetry and chiral limit for the $u,d$
quarks in the correlators, so that the sum rules for the charged and neutral $D_0^*$ mesons coincide. We however retain the $s$-quark mass throughout our computations, hence the violation of $SU(3)_{fl}$ symmetry is to a large extent taken into account.

In the remainder of this section, we derive the two-point QCD sum rules for the decay constants $f_{D^{*(\prime)}_{0}}$ and $f_{D^{*}_{s0}}$, which are defined as
\begin{align}
\label{eq:fD0J}
    &
    \langle 0 |J| D_{0}^{*(\prime)}(p)\rangle =m^2_{D^{*(\prime)}_{0}} \,f_{D^{*(\prime)}_{0}}\,, 
    &&
    \langle 0 |J_{s} | D_{s0}^{*}(p)\rangle =m^2_{D^{*}_{s0}} \,f_{D^{*}_{s0}}\,.
    &
\end{align}
In \refsubsec{SRhad}, following the scenarios  outlined 
in Table~\ref{tab:spectr}, we consider two different hadronic representations of the correlators \eqref{eq:corr2pt},  while in \refsubsec{SROPE} we calculate the same correlators using an OPE.
Following the usual procedure to derive a QCD sum rule, 
 in \refsubsec{SRres}
we match the hadronic representations with the corresponding OPE expressions and use  the quark-hadron duality
approximation. 
We also perform a numerical analysis of the resulting sum rules and obtain
the values of  the decay constants.

\subsection{Hadronic  representations of the two-point correlator}
\label{sec:SRhad}

Following the two scenarios for the spectrum of the lowest-lying charmed scalar mesons discussed in Section~\ref{sec:reson} and specified in \reftab{spectr}, we consider two different hadronic representation of the correlator (\ref{eq:corr2pt}).

\subsubsection*{Scenario 1}

In this case there is only one resonance, i.e. $D_0^*$, and hence 
the two-point QCD sum rule is derived using the standard procedure of Ref.~\cite{Shifman:1978bx}.
The hadronic dispersion relation for the correlator (\ref{eq:corr2pt}) can be written as
\begin{align}
    \Pi_\HAD(q^2)
    =\int\limits_{0}^\infty ds\, 
    \frac{\rho_{\HAD}(s)}{s - q^2}
    \,,
    \label{eq:HAD2pt}
\end{align}
where subtractions are not shown  for simplicity and the hadronic spectral density $\rho_{\HAD}$ is given by
\begin{align}
    \rho_{\HAD}(s) = f_{D^*_0}^2 m_{D^*_0}^4 \delta(m_{D^*_0}^2 - s)
    +
    \rho_{\cont}(s) \theta(s - s_{\th})
    \,.
\label{eq:rhohad}
\end{align}
Here, $\rho_{\cont}$ is the spectral density of continuum and excited states without the contribution of the 
$D_0^*$ resonance and
$s_{\th} = (m_D + m_\pi)^2$ is the lowest continuum threshold.
The scalar resonances are always located above this threshold and hence there is no gap between them and the continuum hadronic states in this channel.
The situation resembles the light-quark vector and scalar channels where the corresponding resonances ($\rho$ and $f_0$, respectively)  are also located above the two-pion threshold. 
Since here, similar to the conventional QCD sum rules for light mesons, we will apply the quark-hadron duality and replace the integral over $\rho_{\cont}(s)$ by the integrated OPE density, the detailed
structure of $\rho_{\cont}$ plays no role.

We then substitute the spectral density (\ref{eq:rhohad}) 
into the dispersion relation and perform a Borel transform with respect to the variable $q^2$ to remove the subtraction terms and to exponentially suppress the contribution of $\rho_{\cont}$: 
\begin{align}
    \Pi_{\HAD}(M^2)
    & =
    f_{D^*_0}^2 m_{D^*_0}^4 \,
    e^{-m_{D^*_0}^2/M^2}
    + 
    \int\limits_{s_{\th}}^\infty ds\,\rho_{\cont}(s)\,e^{-s/M^2}
    \,.
    \label{eq:HAD2pt1}
\end{align}
The hadronic representation for the correlator $\Pi^s$ can be obtained from the above equation with obvious replacements and taking into account that $s_{\th} = (m_D + m_K)^2$ in this case. Note that a lighter $D_s\pi$ state is decoupled in the isospin symmetry limit.

Furthermore, we  take into account the large total width of the $D^*_0$ meson by replacing
\begin{eqnarray}
e^{-m_{D^*_0}^2/M^2}
\to 
\E(\Gamma_{D^*_0},M^2)
\equiv
\int\limits^{\infty}_{s_{\th}}\!ds\, e^{-s/M^2}\Bigg[
\frac{1}{\pi}
\frac{\sqrt{s}\,\Gamma_{D^*_0}(s)}{(s-m_{D^*_0}^2)^2+s \,\Gamma_{D^*_0}^2 (s)}\Bigg]
\,.
\label{eq:widthfact}
\end{eqnarray}
The energy-dependent width $\Gamma_{D^*_0}(s)$ is defined in terms of the total width of $D^*_0$, assuming that the two-particle 
$D\pi$ state is the dominant final state, hence taking the $S$-wave 
phase-space factor
\begin{eqnarray}
\Gamma_{D_0^*}(s)=\Gamma_{D^*_0}^{\rm tot}
\Bigg[\frac{\lambda^{1/2}(s,m_{D}^2,m_\pi^2)m_{D^*_0}}{
\lambda^{1/2}(m_{D^*_0}^2,m_{D}^2,m_\pi^2)\sqrt{s}}\Bigg]
\,,
\end{eqnarray}
where $\lambda$ is the K\"allen function and $\Gamma_{D^*_0}^{\rm tot}$ is the total width.
In the narrow-width limit, i.e. for $\Gamma_{D^*_0}^{\rm tot}\to 0$, the expression inside square brackets in 
\refeq{widthfact} becomes a  $\delta$-function and $\E(\Gamma_{D^*_0},M^2) = e^{-m_{D^*_0}^2/M^2}$ is restored.
For the strange $D^*_{s0}$ meson, it is sufficient to consider only the 
narrow resonance approximation, since its measured total width 
is very small.

\subsubsection*{Scenario 2}

In this case, we need to disentangle the two resonances, i.e. $D_0^*$ and $D_0^{*\prime}$, to obtain the respective decay constants separately. 
The sum rule for the lighter charmed meson $D_0^*$ is again derived using the standard procedure.
Here, one benefits from the fact that the second resonance $D_0^{*\prime}$ is about $350 \MeV$ heavier.
Thus, a duality interval can be reliably determined, so that  the heavier resonance is considered a part of the spectral density $\rho_{\cont}$. 
This yields again Eqs.~(\ref{eq:HAD2pt})-(\ref{eq:rhohad}),  but with a different numerical value of $m_{D^*_0}$ (as given in \reftab{spectr}) and, correspondingly, with a different spectral density $\rho_{\cont}$.

To derive the sum rule for $D_0^{*\prime}$ we start from the hadronic dispersion relation with isolated contributions of two resonances.
After performing the Borel transform we have 
\begin{align}
    \Pi_{\HAD}(M^2)
    & =
    f_{D^*_0}^2 m_{D^*_0}^4 
    e^{-m_{D^*_0}^2/M^2} 
    + 
    f_{D^{*\prime}_0}^2 m_{D^{*\prime}_0}^4 
    e^{-m_{D^{*\prime}_0}^2/M^2}
    + 
    \int\limits_{s_{\th}}^\infty ds\,\tilde{\rho}_{\cont}(s)\,e^{-s/M^2}
    \,,
    \label{eq:HAD2pt2}
\end{align}  
where $\tilde{\rho}_{\cont}(s) $ is now
defined as the spectral density of continuum and excited states
without the contributions of the two lower resonances.

We then follow Ref.~\cite{Gelhausen:2014jea}, 
    where the sum rules for radially excited heavy-light mesons were obtained, and eliminate the contribution of the lighter resonance in the above relation by applying the operator
\begin{align}
    \label{eq:opD}
    \D \equiv \frac{d}{d(1/M^2)}+m_{D_0^*}^2\,
\end{align}
which yields 
\begin{align}
    \D \Pi_{\HAD}(M^2) 
    = 
    f_{D^{*\prime}_0}^2
    (m_{D_0^*}^2 - m_{D_0^{*\prime}}^2)
    m_{D^{*\prime}_0}^4 
    e^{-m_{D^{*\prime}_0}^2/M^2}
    + 
    \int\limits_{s_{\th}}^\infty ds\,(m_{D_0^*}^2 - s)\tilde{\rho}_{\cont}(s)\,e^{-s/M^2}
    \,.
\label{eq:Dpihad}
\end{align}
This procedure enables us to derive the sum rule for the decay constant of the second resonance. 
In this scenario, we neglect the meson widths for simplicity. 
For charmed-strange mesons, we postpone the two-resonance  scenario, until there is  evidence for  the second charmed-strange     scalar resonance.

\subsection{OPE of the two-point correlators}
\label{sec:SROPE}

In the spacelike region, $q^2\ll m_c^2$, we calculate the correlator \eqref{eq:corr2pt}, applying the OPE in local operators.
The result can be written as a truncated series of vacuum averaged  operators of increasing dimension $d$.
Their corresponding Wilson coefficients depend on the choice of  interpolating currents
and encode the short-distance propagation of virtual quarks in the correlator.
The leading power contribution is  given by the  $d=0$ unit operator multiplied by the corresponding Wilson coefficient and represents a purely 
perturbative contribution to the correlator \eqref{eq:corr2pt}.
The contributions of higher dimensional operators, 
starting from $d=3$, are reduced to vacuum condensates which encode the non-perturbative QCD effects in a universal way. 
Their Wilson coefficients are power suppressed, and hence the series can be safely truncated.
For convenience, we separate the perturbative contribution from the condensate part of the OPE:
\begin{align}
    \Pi_{\OPE} (q^2) = 
    \Pi_{\pert} (q^2) + \Pi_{\cond} (q^2)
    \,.
\label{eq:Pi_split}
\end{align}
To use the quark-hadron duality approximation, we replace the 
perturbative part by its dispersion representation, similar to the one in \refeq{HAD2pt}.
The corresponding spectral density at leading order in $\alpha_s$ reads
\begin{align}
    \rho_{\pert}^\LO(s) =
    \frac{3}{8 \pi^2  s}
     (m_c-m_d)^2 \left(s-(m_c+m_d)^2 \right)  \lambda
   ^{1/2}\left(s,m_c^2,m_d^2\right) \theta \left(s-(m_c+m_d)^2\right)   \,.
   \nonumber
\end{align}
Here, the light-quark mass dependence,
which is important for the $s$-quark case,
is exact as opposed to the expanded expressions in $m_s$ that are
given usually in the literature.
For the next-to-leading order  correction of $O(\alpha_s)$  it is possible to  use previous calculations of this correlator from the literature, see e.g., Ref.~\cite{Jamin:1992se}. We find it more convenient and straightforward to use the formula from Ref.~\cite{Gelhausen:2013wia} (see Eq.~(A.2) therein), where 
the correlator of two pseudoscalar heavy-light currents was calculated. 
Performing
the replacement $m_d \to -m_d$, we adapt this formula to our case
of scalar currents.

In the same way, using Eqs.~(B.6),(B.14) and (B.15) 
of Ref.~\cite{Gelhausen:2013wia},
we obtain the condensate part of \refeq{Pi_split}, including the $d=3$ quark-condensate contribution (up to  $O(\alpha_s)$), 
as well as the gluon, quark-gluon and four-quark condensate contributions with $d=4,5$ and $6$, respectively. All these formulas have been  obtained for a nonvanishing light-quark mass.
After Borel transform, the OPE (\ref{eq:Pi_split}) for the correlator in the adopted approximation
becomes 
\begin{equation}
\Pi_{\OPE}(M^2) = 
\int\limits_0^{\infty} ds\, e^{-s/M^2} 
\left[ \rho_{\pert}^\LO(s) +\frac{\alpha_s}{\pi}\rho_{\pert}^\NLO(s)\right]\
+ \Pi_{\cond}(M^2)\,.
\label{eq:PiopeM}
\end{equation}
The analogous result for the correlator $\Pi^s$ can be achieved with obvious replacements in the formulas discussed above.

\subsection{Sum rules and numerical results for decay constants}
\label{sec:SRres}

To finally obtain the two-point QCD sum rule,  
we equate the two Borel transformed analytical expressions for the correlator: the hadronic dispersion relation  (\ref{eq:HAD2pt}) and the result of the OPE  calculation in \refeq{PiopeM}.
Semi-global quark-hadron duality is then used to remove the contributions of the continuum and excited states in the hadronic part of this equation.
This assumption consists in equating the integrated spectral density $\rho_{\rm cont}$ in \refeq{HAD2pt}  to the integrated OPE spectral density:
\begin{equation}
\int\limits_{s_{\th}}^{\infty} d s \,\rho_{\cont}(s)e^{-s/M^2} 
=
\int\limits_{s_0}^{\infty} d s\, \rho_{\OPE}(s) e^{-s/M^2}
\,,
\end{equation}
where $s_0$ is an effective threshold, not necessarily equal to $s_{\th}$. 

Applying this procedure, we obtain the sum rules for the decay constants defined in \refeq{fD0J} in the scenario 1:
\begin{align}
\label{eq:2ptsum1}
    f_{D^*_0}^2 m_{D^*_0}^4 
    \E(\Gamma_{D^*_0},M^2,s_0)
    & =
    \Pi_\OPE(M^2,s_0)
    \,,
\end{align}
and in the scenario 2:
\begin{align}
\label{eq:2ptsum2}
    f_{D^*_0}^2 m_{D^*_0}^4 
    e^{-m_{D^*_0}^2/M^2}
    & =
    \Pi_\OPE(M^2,s_0)
    \,,
    \\
\label{eq:2ptsum3}
    f_{D^{*\prime}_0}^2
    (m_{D_0^*}^2 - m_{D_0^{*\prime}}^2)
    m_{D^{*\prime}_0}^4 
    e^{-m_{D^{*\prime}_0}^2/M^2}
    & =
    \D \Pi_\OPE(M^2,s_0')    
    \,.
\end{align}
The functions $\E(\Gamma_{D^*_0},M^2,s_0)$ and $\Pi_\OPE(M^2,s_0)$ are equal to those defined  in Eqs.~(\ref{eq:widthfact}) and (\ref{eq:PiopeM}), respectively, but with the upper limits of integrals taken at $s_0$, reflecting the use of quark-hadron duality.

\begin{table}[t!]
\renewcommand{\arraystretch}{1.2}
\centering
\begin{tabular}{|c|c|c|}    
\hline
Parameter & Central value $\pm$ Uncertainty/Interval & Ref.
\\\hline\hline
normalisation scale  & $\mu= [1.3,2.5]$ GeV& 
\cite{Gelhausen:2014jea, Khodjamirian:2020mlb}
\\
$s$-quark mass & $m_s(\mu=2 \GeV)=93.4_{+8.6}^{-3.4}$ MeV & \cite{ParticleDataGroup:2022pth}
\\
$c$-quark mass & $m_c(\mu=1.5 \GeV)=1.205\pm 0.035$ GeV & \cite{ParticleDataGroup:2022pth}
\\
strong coupling & $\alpha_s(\mu=1.5 \GeV)=0.353\pm 0.006$ & \cite{ParticleDataGroup:2022pth}
\\
quark condensate  & 
$\langle \bar q q \rangle(\mu=1.5 \GeV)=
-\left(0.278\pm 0.022\; \mbox{GeV} \right)^3$& 
\cite{FlavourLatticeAveragingGroupFLAG:2021npn}
\\
$s$-quark condensate  & 
$
\langle \bar s s \rangle /
\langle \bar q q \rangle= 0.8\pm 0.3$& 
\cite{Ioffe:2002ee}
\\
ratio $\langle \bar{q}Gq \rangle/\langle \bar{q}q \rangle$ &
$m_0^2=0.8\pm 0.2$ GeV$^2$ &
\cite{Ioffe:2002ee}\\
gluon condensate  &$\langle  GG \rangle= 0.012^{+0.006}_{-0.012}\,\mbox{GeV}^4$ &\cite{Ioffe:2002ee}\\
four-quark condensate  & $r_{vac}= 0.5$ &\cite{Ioffe:2002ee}\\
\hline
\end{tabular}
\caption{\emph{Numerical inputs used for the decay constants calculation. 
}
}
\label{tab:inputs2pt}
\end{table}

All the input parameters needed for the numerical evaluation of these sum rules are listed in \reftab{inputs2pt} except for the Borel parameter $M^2$ and the effective thresholds $s_0^{(\prime)}$, which require a dedicated discussion.
On the one hand, the Borel parameter has to be chosen large enough, such that higher power corrections in the OPE are sufficiently suppressed.
On the other hand, the same parameter has to be chosen such small enough, that the contribution of continuum and excited states is subleading compared to the one of the lowest resonance.
We found that these two requirements are satisfied in the range
\begin{align}
    \label{eq:BorelW}
    M^2=[3.0,4.0]\,\GeV^2
    \,,
\end{align}
and also in this range the sum rules have a mild dependence on $M^2$. 
The resulting variation is included in our uncertainty estimate.

The effective threshold $s_0^{(\prime)}$ is determined using the following standard procedure: we take the derivative of each sum rule \eqref{eq:2ptsum1}-\eqref{eq:2ptsum3} 
with respect to $-1/M^2$ 
and divide the  result by the corresponding initial sum rule, obtaining:
\begin{align}
    \label{eq:daughterSR1}
    m_{D^*_0}^2
    & =
    \frac{
        \frac{d}{d(-1/M^2)}
        \Pi_\OPE(M^2,s_0)
        }{
        \Pi_\OPE(M^2,s_0)
        }
    \,,
    \\
    \label{eq:daughterSR2}
    m_{D^{*\prime}_0}^2
    & =
    \frac{
        \frac{d}{d(-1/M^2)}
        \D\Pi_\OPE(M^2,s_0^{\prime})
        }{
        \D\Pi_\OPE(M^2,s_0^{\prime})
        }
    \,.
\end{align}
The effective threshold is then fixed by demanding that these constraints are fulfilled. 
In this procedure, we neglect the width of $D^*_0$, hence \refeq{daughterSR1}
 for the lowest scalar resonance holds for both scenarios. 

Using the $D_0^*$ and $D^{*\prime}_0$ masses in \reftab{spectr}, the inputs in \reftab{inputs2pt} and the Borel window in \refeq{BorelW} we find, for scenario 1,
\begin{align}
    s_0 = 6.84 \pm 0.89 \GeV^2 \,,
\end{align}
and, for scenario 2,
\begin{align}
    s_0 & = 5.50 \pm 0.72 \GeV^2 \,, \\
    s_0' & = 7.80 \pm 0.51 \GeV^2\,.
\end{align}
We can now obtain the numerical values for the decay constants.
Evaluating the sum rule (\ref{eq:2ptsum1}) (scenario 1) we find
\begin{align}
    f_{D^*_0} = 182 \pm 22 \MeV \,,
\label{eq:fd0sc1}
\end{align}
while the corresponding sum rules (\ref{eq:2ptsum2})-(\ref{eq:2ptsum3}) (scenario 2) 
yield
\begin{align}
    f_{D^*_0} & = 146 \pm 17 \MeV \,, \\
    f_{D^{*\prime}_0} & = 170 \pm 38 \MeV\,,
\end{align}
respectively.
The quoted uncertainties are parametric and are obtained varying all input
parameters independently.
We can compare our result (\ref{eq:fd0sc1}) for scenario 1 
with a considerably earlier two-point sum rule prediction in Ref.~\cite{Colangelo:1991ug},
$ m_{D^*_0} = 2.5 \pm 0.1 \GeV $, $ f_{D^*_0} = 170 \pm 20 \MeV $ which is in the same
ballpark.\footnote{
    One should keep in mind that the input parameters including the $c$-quark mass were at that time different.
}

We also perform a numerical analysis for the charmed strange meson ${D^*_{s0}}$
which is well identified independent of the scenario adopted for its nonstrange 
counterparts. 
Adapting the sum rule (\ref{eq:2ptsum1}) to the ${D^*_{s0}}$ meson case, we obtain
\begin{align}
    s_0 = 6.50 &\pm 0.83 \GeV^2 \,, \\
    f_{D^*_{s0}} = 123 &\pm 21 \MeV \,.
\end{align}
Note that from an earlier sum rule calculation of this decay constant 
in Ref.~\cite{Colangelo:2005hv} a similar result was obtained, albeit with small differences in the input parameters.
Also in that paper, a different charmed-strange scalar current is used, without a quark mass prefactor, and hence the result should be  scale-dependent. 

Finally, let us comment on the difference between the decay constants of charmed-strange and charmed-nonstrange 
scalar mesons. 
The fact that the former decay constant is smaller than the latter, opposite to the case of
 charmed pseudoscalar mesons (see e.g.\cite{Gelhausen:2013wia}), partly reflects the different relative sign of the 
$s$-quark mass in the prefactor of the interpolating current and in the OPE formulas. 
In addition, 
we notice that the difference between $f_{D^*_{0}}$ in scenario 1 and $f_{D^*_{s0}}$ is unnaturally large for a typical $SU(3)$ flavour violation. 
A  part of this effect
is due to the effective increase of  
duality threshold by a relatively large $D_0^*$ mass used as an input. 
Additional increase of $f_{D^*_{s0}}$ is due to the fact that 
in scenario 1 the $D_0^*$ is a broad state and hence the sum rule contains  
the width factor (\ref{eq:widthfact})  which --- given the large width
of $D_0^*$ --- is smaller than a simple Borel exponent for a narrow resonance. 
In scenario 2, on the contrary,
the  $SU(3)$ flavour symmetry is violated within the expected $15 - 20\%$.
In this respect, this scenario
in which, as we already discussed, the lighter $D_0^*$ state seems to be a more natural $SU(3)$ partner of $D_{s0}$, seems more plausible.

\section{
    \texorpdfstring{LCSRs for the $\boldsymbol{B_{(s)} \to D_{(s)0}^*}$ form factors}{}
}
\label{sec:lcsr}

We obtain the LCSRs for the $B \to D_{0}^*$ 
and  $B_{s} \to D_{s0}^*$
form factors from the following correlator
which is a time-ordered product of the two currents sandwiched between the $B_{(s)}$ and vacuum states:
\begin{align}
\label{eq:corrLCSR}
\F_{\mu}(p, q) 
&= 
i \!\int \!d^4 x\, e^{ip\cdot x}\, \langle 0|
\T \left\{ J_{(s)}^\dagger(x), J^{\rm A}_\mu(0) \right\} | \bar{B}_{(s)}(p+q)\rangle
\nonumber\\
&\equiv
i\,p_\mu\, \F^{(p)}(p^2,q^2)
+i\,q_\mu\,\F^{(q)}(p^2,q^2)\,,
\end{align}
where $J_{(s)}^\dagger$ is the conjugate of the scalar current (\ref{eq:J}) with four-momentum $p$ and  $J^{\rm A}_\mu=\bar{c}\gamma_\mu\gamma_5 b$ is the axial part of the weak current with four-momentum $q$. The vector-current part of the transition matrix element between $B$ and a scalar meson vanishes due to the $P$-parity conservation. With our 
four-momenta assignment, the momentum of the $B$-meson state is $p+q$, so that $(p+q)^2=m_B^2$. The r.h.s. of \refeq{corrLCSR} contains a decomposition
of the correlation function in two  
Lorentz structures which we have simply chosen 
to be equal to the two independent four-momenta.

In this section, for definiteness we obtain LCSR for the $\bar{B}_0\to D_0^{*+}$ form factors.
The results for the transition of $B^-$ into a neutral  $D_0^*$ 
are the same in the isospin symmetry limit. The LCSRs for 
$\bar{B}_s\to D^+_{s0}$ form factors are obtained by 
replacing the non-strange heavy mesons by their strange counterparts.
As we shall see, in the adopted approximation for LCSRs, we have
to  replace the meson masses and the parameters of 
$B$-meson DAs, and all of them only implicitly depend on $m_s$.
Explicitly, the $s$-quark mass only enters the normalisation factor
of the charmed scalar current. 

In the rest of this section, we present the hadronic dispersion relation for the correlator \eqref{eq:corrLCSR} in \refsubsec{LChad}, while we perform the OPE calculation of the same correlator in \refsubsec{LCOPE}.
We derive the LCSRs and obtain the corresponding numerical results in \refsubsec{LCSRanalytic}.

\subsection{Hadronic dispersion relation}
\label{sec:LChad}

The hadronic matrix element of the $B\to D_0^*$ transition 
is decomposed into two form factors in two different ways:
\begin{align}
\label{eq:BD0stff}
\langle D^*_0(p)| J^{\rm A}_\mu| \bar{B}(p+q)\rangle
&=i(2p_\mu+q_\mu)f^{BD^*_0}_+(q^2)
+iq_\mu f^{BD^*_0}_-(q^2)
\nonumber\\[2mm]
&=i\!\left(2p_\mu-\frac{m_B^2-m^2_{D_0^*}-q^2}{q^2} q_\mu \right) f^{BD^*_0}_+(q^2)
+i\frac{m_B^2-m^2_{D_0^*}}{q^2}q_\mu f^{BD^*_0}_0(q^2)\,,
\end{align}
where  $f^{BD^*_0}_0$ is related 
to $f^{BD^*_0}_\pm$ via:
\begin{equation}
\label{eq:rel}
f^{BD^*_0}_0(q^2)=f^{BD^*_0}_+(q^2) + \frac{q^2}{m_B^2-m_{D_0^*}^2}f^{BD^*_0}_-(q^2)\,,
\end{equation}
so that $f^{BD^*_0}_0(0)=f^{BD^*_0}_+(0)$.
These definitions and relations are analogous to the standard ones for the $B\to D$ vector and scalar 
form factors of the vector weak current. 
In our case, the vector current is replaced by the 
axial one and the pseudoscalar meson is replaced by the 
scalar one. Similarly, the form factor $f^{BD^*_0}_0(q^2)$ can be expressed as the $B\to D^*_0$ matrix element of the pseudoscalar current:
\begin{equation}
\label{eq:rel2}
f^{BD^*_0}_0(q^2)= \frac{m_b+m_c}{m_B^2-m_{D_0^*}^2}
\langle D^*_0(p)|\bar{c}i\gamma_5 b | \bar{B}(p+q)\rangle\,.
\end{equation}

\subsubsection*{Scenario 1}

To obtain the hadronic dispersion relation for the correlation function (\ref{eq:corrLCSR}), we write down the imaginary part of it
in the variable $p^2=s$ for $s>0$ and fixed $q^2$: 
\begin{align}
\label{eq:had_Imp}
\mbox{Im}_{p^2}\F_{\mu}(p, q)= 
\pi\delta(s-m^2_{D_0^*})
\langle 0| J^\dagger|D_0^*(p)
\rangle \langle D_0^*(p) J^{\rm A}_\mu | \bar{B}(p+q)\rangle + \pi \rhomu(p,q) \theta(s - s_{\th}) 
\,.
\end{align}
As discussed in \refsubsec{SRhad}, $s_{\th}=(m_D+m_\pi)^2$ and 
we isolate the lowest $D_0^*$ pole (temporarily neglecting its total width). We also denote by
$\rhomu$ the spectral density of continuum and excited states
with the $D_0^*$ quantum numbers.
For this spectral density we  use  the
same decomposition in invariant functions as in \refeq{corrLCSR}: 
$$\rhomu (p,q)= p_\mu \,\rhop(s,q^2)+q_\mu \,\rhoq(s,q^2)\,.$$
Substituting the decomposition (\ref{eq:BD0stff}) and 
the decay constant definition  (\ref{eq:fD0J}) into \refeq{had_Imp}
and separating the two kinematical structures according to 
\refeq{corrLCSR}, we  obtain  hadronic dispersion 
relations for the two invariant amplitudes:
\begin{align}
\Fp(p^2,q^2)&= \frac{1}{\pi}\int\limits_{s_{\th}}^{\infty}
ds\frac {\mbox{Im}_{p^2}\Fp(s, q^2)}{s-p^2}=
\frac{2 m_{D_0^*}^2 f_{D_0^*}f_+^{BD^*_0}(q^2)}{m_{D_0^*}^2-p^2}+
\int\limits_{s_{\th}}^{\infty}
ds\ \frac {\rhop(s, q^2)}{s-p^2}\,,
\label{eq:dispLCSR1}
\\
\Fq(p^2,q^2)&= \frac{1}{\pi}\int\limits_0^{\infty}
ds\, \frac{\mbox{Im}_{p^2}{\Fq}(s, q^2)}{s-p^2}=
\frac{ m_{D_0^*}^2 f_{D_0^*}\big[f_+^{BD^*_0}(q^2)+f_-^{BD^*_0}(q^2)\big]}{m_{D_0^*}^2-p^2}+
\int\limits_{s_{\th}}^{\infty}
ds\ \frac{\rhoq(s, q^2)}{s-p^2}\,.
\label{eq:dispLCSR2}
\end{align}
Their Borel transform yields
\begin{align}
\Fp(\hat{M}^2,q^2)&= 
2 m_{D_0^*}^2 f_{D_0^*}f_+^{BD^*_0}(q^2)\,e^{-m_{D^*_0}^2/M^2}
+ \int\limits_{s_{\th}}^{\infty}
ds\ \rhop(s, q^2)\, e^{-s/M^2} \,,
\label{eq:dispM2LCSR1}
\\
\Fq(\hat{M}^2,q^2)&= 
m_{D_0^*}^2 f_{D_0^*}\big[f_+^{BD^*_0}(q^2)+f_-^{BD^*_0}(q^2)\big]\,e^{-m_{D^*_0}^2/M^2}+
\int\limits_{s_{\th}}^{\infty}
ds\ \rhoq(s, q^2)\, e^{-s/M^2}\,.
\label{eq:dispM2LCSR2}
\end{align}
Similar to the case of two-point sum rules (see \refsubsec{SRhad}), we  improve the accuracy of these relations by introducing
the $s$-dependent total width of $D_0^*$, i.e., by performing the replacement \eqref{eq:widthfact}.

\subsubsection*{Scenario 2}

In scenario 2, the hadronic dispersion relation $\Fi$ for the lowest-lying scalar resonance has the same  form of \refeqs{dispM2LCSR1}{dispM2LCSR2} but with different mass and decay constant of $D_0^*$.
The second scalar resonance $D_0^{*\prime}$ can be isolated using again the operator \eqref{eq:opD}, which yields
\begin{align}
\D\Fp(\hat{M}^2,q^2)&= 
2 (m_{D_0^*}^2 - m_{D_0^{*\prime}}^2)m_{D_0^{*\prime}}^2 f_{D_0^{*\prime}}f_+^{BD^{*\prime}_0}(q^2)\,e^{-m_{D^{*\prime}_0}^2/M^2}
\nonumber\\*
&+ \int\limits_{s_{\th}}^{\infty}
ds\ (m_{D_0^*}^2 - s)\rhop(s, q^2)\, e^{-s/M^2} \,,
\label{eq:dispM2s2LCSR1}
\\
\D\Fq(\hat{M}^2,q^2)&= 
(m_{D_0^*}^2 - m_{D_0^{*\prime}}^2)m_{D_0^{*\prime}}^2 f_{D_0^{*\prime}}\big[f_+^{BD^{*\prime}_0}(q^2)+f_-^{BD^{*\prime}_0}(q^2)\big]\,e^{-m_{D^{*\prime}_0}^2/M^2}
\nonumber\\*
&+\int\limits_{s_{\th}}^{\infty}
ds\ (m_{D_0^*}^2 - s)\rhoq(s, q^2)\, e^{-s/M^2}\,,
\label{eq:dispM2s2LCSR2}
\end{align}
in analogy with \refeq{Dpihad} for the two-point sum rule.

\subsection{Light-cone OPE for the correlator}
\label{sec:LCOPE}

\begin{figure}[t!]
\centering
\includegraphics[scale=0.45]{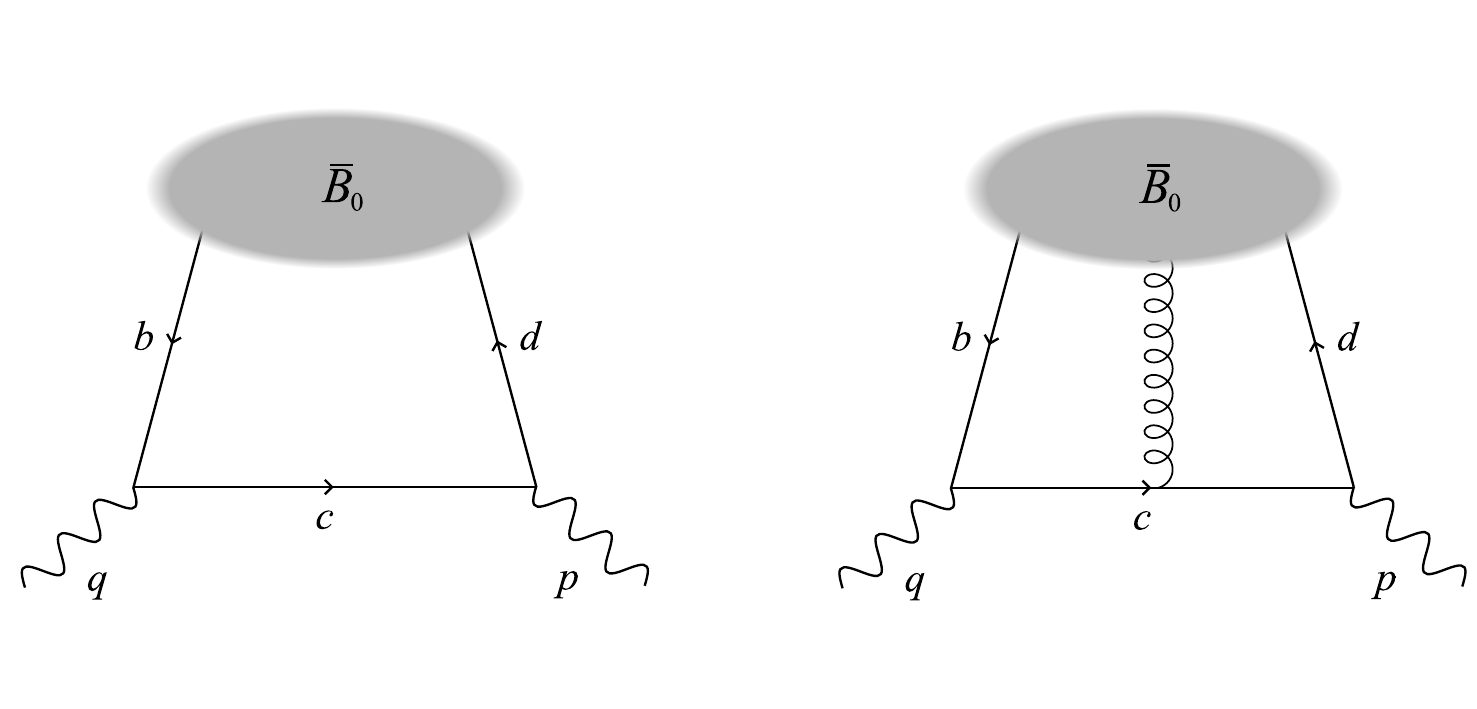}
	\caption{\emph{The leading order (left panel) and  soft gluon emission (right panel) diagrams for the correlation function (\ref{eq:corrLCSR}).}}
	\label{fig:diagLO}  
\end{figure} 

For $p^2\ll m_c^2$ and $q^2\ll (m_b+m_c)^2$, that is  far below the hadronic thresholds in the channels of the
interpolating and weak currents, 
the correlator (\ref{eq:corrLCSR}) can be calculated, expanding the time-ordered
product of currents near the light-cone $x^2\simeq 0$~\cite{Khodjamirian:2006st}. 
In practical terms, we 
have to compute the diagrams depicted in~\reffig{diagLO}. 
At leading order in $\alpha_s$, the diagram on the left panel 
consists of the virtual $c$-quark 
propagating between the two vertices, whereas 
a quark-antiquark pair emitted at a light-cone separation forms a $B$-meson state.
This long-distance part of the diagram is encoded in terms of the two-particle $B$-meson DAs
defined in  HQET. The computation of this diagram is described in detail in 
Ref.~\cite{Gubernari:2022hrq}. 
The only difference in our case is that 
the Dirac-structure in the vertex of the interpolation current has changed 
to a unit matrix. 

In addition, to improve the accuracy of the light-cone OPE,
we also take  into account the effect of  soft (low virtuality) gluon 
emitted from the $c$-quark line 
and absorbed, together with the quark-antiquark pair, in the three-particle 
$B$-meson DAs. The diagram is shown on the right panel of Fig.~\ref{fig:diagLO}.
To evaluate this diagram we need the one-gluon term in the light-cone expansion of the $c$-quark propagator \cite{Balitsky:1987bk}. For the latter,
we use the most convenient symmetric form: 
\begin{equation}
\langle 0 | T\{c(x)\bar{c}(0)\} |0 \rangle _G=
-i\!\int\limits_0^1\!du\,G_{\mu\nu}(ux)
\!\int\!\frac{d^4f}{(2\pi)^4}e^{-ifx}                                \frac{\bar{u}(\slash\!\!\!\!f +m_c)\sigma^{\mu\nu}
+u\sigma^{\mu\nu}(\slash\!\!\!\!f +m_c)}{2(f^2-m_c^2)^2}          \,, \label{eq:quarkpropG}
\end{equation}
where $G_{\mu\nu}=g_sG^a_{\mu\nu}(\lambda^a/2)$.
This diagram, for a similar correlator, but with 
 a different interpolating current,
was computed  earlier in Ref.~\cite{Faller:2008tr}, where the LCSRs for the $B\to D^{(*)}$
form factors were obtained. In Ref.~\cite{Gubernari:2018wyi},  
the contributions of three-particle DAs to this sum rule were improved, 
taking into account a more complete set of three-particle $B$-meson DAs from Ref.~\cite{Braun:2017liq} which we also use here. 
We will not present further
 details of our computation, referring e.g., to Ref.~\cite{Descotes-Genon:2019bud} where the contributions of three-particle DAs are also obtained and described for a  similar correlator, but with the virtual $s$ quark. 

The OPE result for the invariant amplitudes $\Fi$ with $i=p,q$ can be reduced to the following generic form:
\begin{equation}
\Fi_{\OPE}(p^2,q^2)
=(m_c-m_d)\,m_Bf_B\sum_{k=1}^{4} \int\limits_0^\infty d\sigma \frac{I^{(i,k)}(\sigma,q^2)}{(p^2-s(\sigma,q^2))^k}\,,
\label{eq:Fexpan}
\end{equation}
where the variable 
\begin{equation}
s(\sigma,q^2)=\sigma m_B^2-\frac{\sigma q^2-m_c^2}{1-\sigma}\,,
\label{eq:s}
\end{equation}
is introduced, so that, inversely,
\begin{equation}
\sigma(s,q^2)=\frac{m_B^2-q^2+s-\sqrt{4 \left(m_c^2-s\right)m_B^2+\left(m_B^2-q^2+s\right)^2}}{2 m_B^2} \label{eq:sig_s}\,.  
\end{equation}
The functions $I^{(i,k)}$ are defined as:
\begin{equation}
I^{(i,k)}(\sigma,q^2)=\sum_\psi 
\C_\psi^{(i,k)}(\sigma,q^2)\,\psi(\sigma m_B)
+\sum_{\chi_j} 
\!\!\int\limits_0^{\sigma m_B}\!\! d\omega_1\!\!\int\limits_{\sigma m_B-\omega_1}^\infty\!\!
\frac{d\omega_2}{\omega_2}\,\,
\C_{\chi_j}^{(i,k)}(\sigma,\omega_1,\omega_2,q^2) \,
\chi_i(\omega_1,\omega_2)
\,,
\label{eq:Iexpan}
\end{equation}
where the first sum goes over the contributions of the four two-particle $B$-meson DAs:
$$
\psi(\omega)=\Big\{ \phi_+(\omega),~\bar{\Phi}_{\pm}(\omega), 
~g_+(\omega),~\overline{G}_{\pm}(\omega)\Big\}\,,
$$
and the second sum contains contributions of the eight linear combinations of three-particle  DAs: 
\begin{align*}
    \chi_j(\omega_1,\omega_2)=\Big\{\chi_1(\omega_1,\omega_2), \dots, \chi_8(\omega_1,\omega_2)\Big\}\,.
\end{align*}
To derive \refeq{Fexpan} we have performed the replacement $\omega \mapsto \sigma m_B$ and $ u \mapsto (\sigma\,m_B - \omega_1)/\omega_2$ for the coefficients of the two- and three-particle terms, respectively. 
The definitions of  DAs and of their combinations are given in  Appendix~\ref{app:DA}, where we also specify
their model and its parameters.
We recall that the $B$-meson DAs are defined in HQET, i.e. in the limit $m_B\to \infty$, and the variables $\omega$ and $\omega_{1,2}$ are understood as the plus components 
of the momenta of light degrees of freedom inside $B$.
This explains a formally infinite upper limit in \refeq{Fexpan}. 
In reality, these variables are limited from above by $O(m_B)$, hence, the integration is supported only in the interval $0<\omega<1$.
This is consistent with the exponential falloff of the model DAs adopted here and presented in Appendix
 \ref{app:DA}.
The OPE in \refeq{Fexpan} is determined by the coefficients $\C_\psi^{(i,k)}$ and $\C_{\chi_j}^{(i,k)}$
entering \refeq{Iexpan}. 
Our main analytical result are the expressions for these coefficients 
collected in Appendix~\ref{app:OPE}.
\\

In order to use quark-hadron duality in the derivation of  LCSRs, we need to recast the OPE (\ref{eq:Fexpan}) in the form of a dispersion integral in the variable $p^2$.
In addition, we perform the Borel transform with respect to $p^2$.  
Following Ref.~\cite{Gubernari:2018wyi}, the OPE result can be written as
\begin{align}
    \Fi_{\OPE}(\hat{M}^2, q^2)
    &=
    \Fi_{\OPE}(\hat{M}^2, \hat{s}_0, q^2)
    +
    \Fi_{\OPE}(\hat{M}^2, \infty , q^2)
    \,,
    \label{eq:MasterFor1}
\end{align}
where
\begin{align}
\label{eq:Fs0}
    \Fi_{\OPE}(\hat{M}^2, \hat{s}_0, q^2)
    &=
   (m_c-m_d)m_B f_B  \sum_{k=1}^{4}
    \frac{(-1)^{k}}{(k-1)!}
    \Bigg\{\int\limits_0^{\sigma_0} d \sigma \;e^{-s(\sigma,q^2)/\hat{M}^2} \frac{1}{(\hat{M}^2)^{k-1}}
    I^{(i,k)}(\sigma,q^2)
    \nonumber\\*
    & + \Bigg[e^{-s(\sigma,q^2)/\hat{M}^2}\sum_{j=1}^{k-1}\frac{1}{(\hat{M}^2)^{k-j-1}}\frac{1}{s'}
    \left(\frac{d}{d\sigma}\frac{1}{s'}\right)^{j-1}
    I^{(i,k)}(\sigma,q^2)
    \Bigg]_{\sigma=\sigma_0} \Bigg\}
\end{align}
and
\begin{align}
\label{eq:Finfty}
    \Fi_{\OPE}(\hat{M}^2, \infty , q^2)
    &
    = (m_c-m_d)m_B f_B \sum_{k=1}^{4}
    \frac{(-1)^{k}}{(k-1)!}
    \int\limits_{\sigma_0}^\infty d \sigma \;
    e^{-s(\sigma,q^2)/\hat{M}^2} 
     \left(\frac{d}{d\sigma}\frac{1}{s'}\right)^{k-1}
     I^{(i,k)}(\sigma,q^2)
    \,.
\end{align}
Here, we have introduced the following notation:
\begin{align*}
&
\left(\frac{d}{d\sigma}\frac{1}{s'}\right)^{n} f(\sigma)
\equiv \left(\frac{d}{d\sigma}\frac{1}{s'}\left(\frac{d}{d\sigma}\frac{1}{s'}\dots f(\sigma)\right)\right) \,, ~~~
&&
s' \equiv \frac{ds}{d\sigma} \,,~~
&&
    \sigma_0 \equiv \sigma(\hat{s}_0,q^2)
    \,.
    &
\end{align*}
The corresponding OPE expressions in the case of $B_s\to D^*_{s0}$ form factors are 
obtained by replacing  $(m_c-m_d)\to (m_c-m_s)$, $f_B\to f_{B_s}$ and $m_B\to m_{B_s}$ in  Eq.~(\ref{eq:Fexpan}).

\subsection{Light-cone sum rules and numerical results for the form factors}
\label{sec:LCSRanalytic}

The LCSRs for $B_{(s)} \to D_{(s)0}^*$ form factors can now be easily obtained by equating the OPE results in \refeq{MasterFor1} with 
the corresponding hadronic representations in \refeqs{dispM2LCSR1}{dispM2s2LCSR2}. 
The semi-global quark-hadron duality approximation is then used to remove the contributions of continuum and excited states. 
In other words we assume that the term denoted as $\Fi_{\OPE}(\hat{M}^2, \infty, q^2)$  in \refeq{MasterFor1} cancels the integral over the spectral density $\rhoi(s, q^2)$.
Following these steps, we derive the sum rules for the form factors in scenario 1:
\begin{align}
2 m_{D_0^*}^2 f_{D_0^*}f_+^{BD^*_0}(q^2)\,
\E(\Gamma_{D^*_0},\hat{M}^2,\hat{s}_0)
&=
\Fp_{\OPE}(\hat{M}^2, \hat{s}_0, q^2)\,,
\label{eq:LCSRs1fpl}
\\
m_{D_0^*}^2 f_{D_0^*}\big[f_+^{BD^*_0}(q^2)+f_-^{BD^*_0}(q^2)\big]
\E(\Gamma_{D^*_0},\hat{M}^2,\hat{s}_0)
&=
\Fq_{\OPE}(\hat{M}^2, \hat{s}_0, q^2)\,,
\label{eq:LCSRs1fpm}
\end{align}
and in scenario 2:
\begin{align}
2 m_{D_0^*}^2 f_{D_0^*}f_+^{BD^*_0}(q^2)\,
e^{- m_{D_0^*}^2/\hat{M}^2}
&=
\Fp_{\OPE}(\hat{M}^2, \hat{s}_0, q^2)\,,
\label{eq:LCSRs2fpl}
\\
m_{D_0^*}^2 f_{D_0^*}\big[f_+^{BD^*_0}(q^2)+f_-^{BD^*_0}(q^2)\big]
e^{- m_{D_0^*}^2/\hat{M}^2}
&=
\Fq_{\OPE}(\hat{M}^2, \hat{s}_0, q^2)\,,
\label{eq:LCSRs2fpm}
\\
2 (m_{D_0^*}^2 - m_{D_0^{*\prime}}^2)m_{D_0^{*\prime}}^2 f_{D_0^{*\prime}}f_+^{BD^{*\prime}_0}(q^2)\,e^{-m_{D^{*\prime}_0}^2/\hat M^2}
&=
\D\Fp_{\OPE}(\hat{M}^2, \hat{s}^\prime_0, q^2)\,,
\label{eq:LCSRs2fplp}
\\
(m_{D_0^*}^2 - m_{D_0^{*\prime}}^2)m_{D_0^{*\prime}}^2 f_{D_0^{*\prime}}\big[f_+^{BD^{*\prime}_0}(q^2)+f_-^{BD^{*\prime}_0}(q^2)\big]\,e^{-m_{D^{*\prime}_0}^2/\hat M^2}
&=
\D\Fq_{\OPE}(\hat{M}^2, \hat{s}^\prime_0, q^2)\,,
\label{eq:LCSRs2fpmp}
\end{align}
\begin{table}[t!]
\renewcommand{\arraystretch}{1.2}
\centering
\begin{tabular}{|c|c|c|}    
\hline
Parameter & Central value $\pm$ Uncertainty & Ref.
\\\hline\hline
$B$-meson decay constant  & 
$f_B=190.0 \pm 1.3$\,MeV  & \cite{FlavourLatticeAveragingGroupFLAG:2021npn} \\
$B_{s}$-meson decay constant  & 
$f_{B_{s}}=230.3 \pm 1.3$\,MeV  & \cite{FlavourLatticeAveragingGroupFLAG:2021npn} \\[2mm]

\multirow{4}{*}{ \parbox{4cm}{\centering Parameters of the \\ $B_{(s)}$-meson DAs}} & 
$\lambda_B=0.460\pm 0.110$\,GeV &
\cite{Braun:2003wx} \\
& $\lambda_{B_s}/\lambda_B=1.19\pm0.14$ &
\cite{Khodjamirian:2020hob} \\
& $\lambda_E^2=0.02\pm 0.03$\,GeV$^2$ & \cite{Nishikawa:2011qk,Rahimi:2020zzo} \\
& $\lambda_H^2=0.11\pm 0.08$\,GeV$^2$ & \cite{Nishikawa:2011qk,Rahimi:2020zzo} \\
\hline
\end{tabular}
\caption{\emph{Numerical inputs used for the form factor calculation from LCSRs. }
    }
\label{tab:inputsLCSR}
\end{table}

The input parameters needed to calculate the form factors  from these sum rules are all listed in \reftab{inputsLCSR}, with the exception of  the quark masses that are in \reftab{inputs2pt} and the decay constants 
of charmless scalar mesons that have been obtained in Section~\ref{sec:2ptSR}. 
In the adopted approximation for the OPE, the $B$-meson DAs entering the LCSRs   are characterised by four parameters (see \refapp{DA} for more details).
One of them is the $B$-meson decay constant for which we adopt the lattice QCD average.  The other three  parameters include 
the first inverse moment $\lambda_B$  and the normalisation parameters $\lambda_E^2$ and $\lambda_H^2$
of three-particle DAs, the latter defined as the vacuum-to-$B$ matrix elements of local quark-antiquark-gluon-operators in HQET \cite{Grozin:1996pq}.
The values of all three parameters in \reftab{inputsLCSR} are taken from the QCD sum rule determinations.
As a conservative choice, the intervals for the parameters $\lambda_E^2$ and $\lambda_H^2$ are obtained by averaging the central values 
obtained from the  two independent sum rules  in Refs.~\cite{Nishikawa:2011qk,Rahimi:2020zzo} and adding their respective uncertainties.
Given the large uncertainties of these two parameters, we take the same intervals also for the $B_s$-meson DAs.
However, for $\lambda_{B_s}$ we use the QCD sum rule estimates from Ref.~\cite{Khodjamirian:2020hob}, which is also consistent with the more recent independent calculation of this parameter in \cite{Feldmann:2023aml}.
Furthermore, we find that the Borel parameter $\hat{M}^2$ of the LCSRs, which is in principle independent from the Borel parameter $M^2$ of the two-point sum rules, can be varied in the same interval of \refeq{BorelW}, i.e. between $3.0\,\GeV^2$ and $4.0\,\GeV^2$.
In fact, we have checked that the LCSRs are stable in this interval and both the duality-estimated contribution of heavier states  and the subleading twist contributions are sufficiently suppressed.
Following Refs.~\cite{Gubernari:2022hrq,Descotes-Genon:2019bud}, we also use for the LCSRs the same threshold as the one determined in \refsubsec{SRres} for the two-point sum rules, that is, we fix $\hat{s}_0^{(\prime)}=s_0^{(\prime)}$. 
\\

For each LCSR considered above, we calculate the 
OPE expression 
$\Fi_\OPE(\hat{M}^2, \hat{s}_0,q^2)$ 
or  $\D \Fi_\OPE(\hat{M}^2, \hat{s}_0,q^2)$, in both cases denoted for brevity as $\Fi_\OPE(q^2)$, where $i=p$ or $q$, at $q^2 = \{-20,\, -15,\allowbreak -10,\, -5\} \GeV^2$, 
varying  all input parameters within their adopted intervals. 
Using the OPE only for negative $q^2$  values is justified by the fact that for $q^2\geq 0$, the  subleading twist contributions become numerically of the same order of magnitude as the leading twist ones at $q^2=0$.
This behaviour is not surprising and has already been noticed in Ref.~\cite{Gubernari:2022hrq}.
We also observe that the contributions of three-particle DAs to LCSRs are   relatively small (of the order of few per cent) compared to the contributions of two-particle DAs, in agreement with the results of Refs.~\cite{Gubernari:2018wyi,Gubernari:2020eft}.

To obtain the form factors in the semileptonic region, that is for
$0<q^2<(m_{B_{(s)}}-m_{D_{(s)0}^*})^2$, 
we fit for each LCSR the OPE results $\F^{(i)}_\OPE (q^2)$ at $q^2\leq 0$ to the following $z$ expansion:
\begin{align}
     \F^{(i)}_{\fit} (q^2)
    =
    \frac{1}{1 - \frac{q^2}{m_{(i)}^2}}
    \sum_{k=0}^K
    \alpha_k^{(i)}
    \left[
        z(q^2)
        -
        z(0)
    \right]^k
    \,.
    \label{eq:zexpOPE}
\end{align}
The $q^2 \mapsto z$ map is defined as
\begin{align}
    \label{eq:zdef}
    z(q^2) = \frac{\sqrt{t_+-q^2} - \sqrt{t_+ - t_0^{\phantom{1}}}}{\sqrt{t_+-q^2} + \sqrt{t_+ - t_0^{\phantom{1}}}}
    \,,
\end{align}
where
\begin{align}
    & t_+ = (m_B + m_D)^2 \,,&
    & t_0 = (m_B + m_D)\left(\sqrt{m_B} - \sqrt{m_D}\right)^2 \,.&
\end{align}
The parameter $m_{(i)}$ in  Eq.~(\ref{eq:zexpOPE}) is the mass of the lightest  $b\bar{c}$ pole  in the timelike region of the form factor related to
$\Fi_\OPE$ via LCSRs. 
In fact, $\F^{(p)}_\OPE$ is associated with the  form factor $f_+$  which contains  $b\bar{c}$ states 
with spin-parity $J^P=1^+$ 
in the timelike region. 
Similarly,   
$\F^{(q)}_\OPE$ is related to a linear combination of $f_+$ and $f_0$, and the latter form factor has lighter states with $J^P=0^-$.
Hence, we take $m_{(p)}=6.767 \GeV$ and $m_{(q)}=6.275 \GeV$, since those are the masses of the lightest $b\bar{c}$ states with $J^P=1^+$ and $J^P=0^-$~ , respectively, as estimated in lattice QCD \cite{Detmold:2015aaa}.

\begin{table}[t!]
\renewcommand{\arraystretch}{1.3}
\centering
\begin{tabular}{|c|c|r|r|c|}    
\hline
Process (scenario) & $(i)$ & $\alpha_0^{(i)} \phantom{00000}$ & $\alpha_1^{(i)} \phantom{00000}$ & Correlation
\\\hline\hline
\multirow{2}{*}{$B\to D_0^*$ (1)} & 
$(p)$ & 
$ -0.100 \pm 0.060 $ &
$ 0.15 \pm 0.15 \phantom{0}$ &
$ -0.93 $ \\
& 
$(q)$ & 
$ 0.019 \pm 0.021 $ &
$ -0.029 \pm 0.085 $ &
$ -0.97 $ \\
\hline
\multirow{2}{*}{$B_s\to D_{s0}^*$ (1)} & 
$(p)$ &
$ -0.060 \pm 0.051 $ &
$ 0.050 \pm 0.16  \phantom{0}$ &
$ -0.97 $ \\
& $(q)$ &
$ 0.004 \pm 0.017 $ &
$ 0.029 \pm 0.076 $ &
$ -0.98 $ \\
\hline
\multirow{2}{*}{$B\to D_0^*$ (2)} & 
$(p)$ &
$ -0.074 \pm 0.053 $ &
$ 0.13 \pm 0.18  \phantom{0}$ &
$ -0.97 $ \\
& $(q)$ &
$ 0.011 \pm 0.018 $ &
$ -0.007 \pm 0.080 $ &
$ -0.98 $ \\
\hline
\multirow{2}{*}{$B\to D_0^{*\prime}$ (2)} & 
$(p)$ &
$ 0.090 \pm 0.075 $ &
$ -0.27 \pm 0.39  \phantom{0}$ &
$ -0.85 $ \\
& $(q)$ &
$ -0.051 \pm 0.024  $ &
$ 0.20 \pm 0.12  \phantom{0}$ &
$ -0.84 $ \\
\hline
\end{tabular}
\caption{\emph{Coefficients of the $z$- expansion~(\ref{eq:zexpOPE}) fitted to  $\F^{(i)}_{\emph{\OPE}}$.
}}
\label{tab:coeffs}
\end{table}

\begin{table}[t!]
\renewcommand{\arraystretch}{1.5}
\centering
\begin{tabular}{|c|c|c|c|c|}  
\hline
Process (scenario) & Form factor & $q^2=0$ & $q^2=\frac{1}{2}q^2_{\rm max}$ & $q^2=q^2_{\rm max}$
\\\hline\hline
\multirow{2}{*}{$B\to D_0^*$ (1)} & 
$f_+^{BD^*_0}$ & 
$ -0.35^{+0.20}_{-0.21} $ &
$ -0.41^{+0.22}_{-0.25} $ &
$ -0.48^{+0.26}_{-0.30} $ \\
& 
$f_-^{BD^*_0}$ & 
$ \phantom{+}0.49^{+0.34}_{-0.37} $ &
$ \phantom{+}0.57^{+0.40}_{-0.45} $ &
$ \phantom{+}0.67^{+0.47}_{-0.56} $ \\
\hline
\multirow{2}{*}{$B_s\to D_{s0}^*$ (1)} & 
$f_+^{BD^*_{s0}}$ &
$ -0.24^{+0.21}_{-0.21} $ &
$ -0.28^{+0.25}_{-0.25} $ &
$ -0.32^{+0.29}_{-0.31} $ \\
& $f_-^{BD^*_{s0}}$ &
$ \phantom{+}0.27^{+0.36}_{-0.37} $ &
$ \phantom{+}0.30^{+0.43}_{-0.45} $ &
$ \phantom{+}0.34^{+0.52}_{-0.56} $ \\
\hline
\multirow{2}{*}{$B\to D_0^*$ (2)} & 
$f_+^{BD^*_0}$ &
$ -0.20^{+0.15}_{-0.14} $ &
$ -0.23^{+0.18}_{-0.17} $ &
$ -0.27^{+0.21}_{-0.21} $ \\
& $f_-^{BD^*_0}$ &
$ \phantom{+}0.25^{+0.25}_{-0.24} $ &
$ \phantom{+}0.29^{+0.30}_{-0.30} $ &
$ \phantom{+}0.34^{+0.36}_{-0.37} $ \\
\hline
\multirow{2}{*}{$B\to D_0^{*\prime}$ (2)} & 
$f_+^{BD^{*\prime}_0}$ &
$ \phantom{+}0.22^{+0.16}_{-0.23} $ &
$ \phantom{+}0.25^{+0.19}_{-0.28} $ &
$ \phantom{+}0.30^{+0.23}_{-0.35} $ \\
& $f_-^{BD^{*\prime}_0}$ &
$ -0.49^{+0.24}_{-0.22} $ &
$ -0.58^{+0.29}_{-0.28} $ &
$ -0.70^{+0.35}_{-0.35} $ \\
\hline
\end{tabular}
\caption{\emph{Form factor values at three selected $q^2$ points.
}}
\label{tab:ffval}
\end{table}

\begin{figure}[p]
\vspace{-2cm}
    \centering
    \newcommand\ww{0.40}
    \begin{tabular}{cc}
        \includegraphics[width=\ww\textwidth]{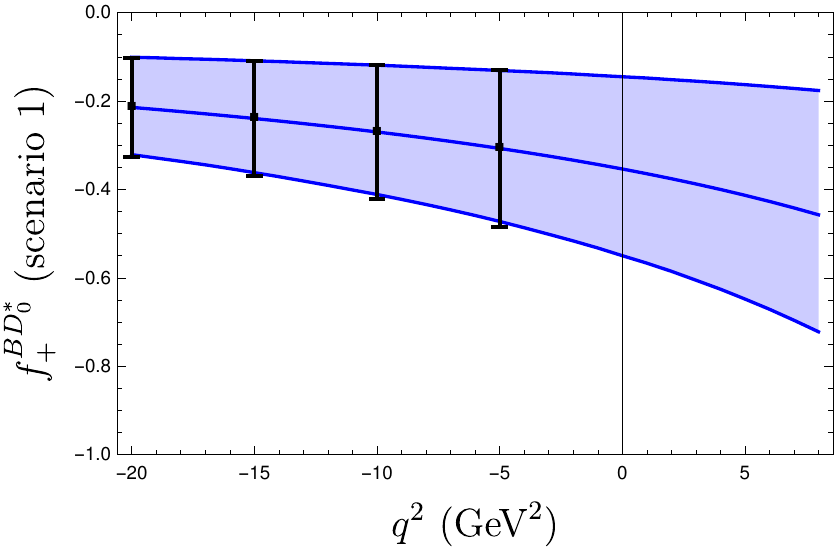}  &
        \includegraphics[width=\ww\textwidth]{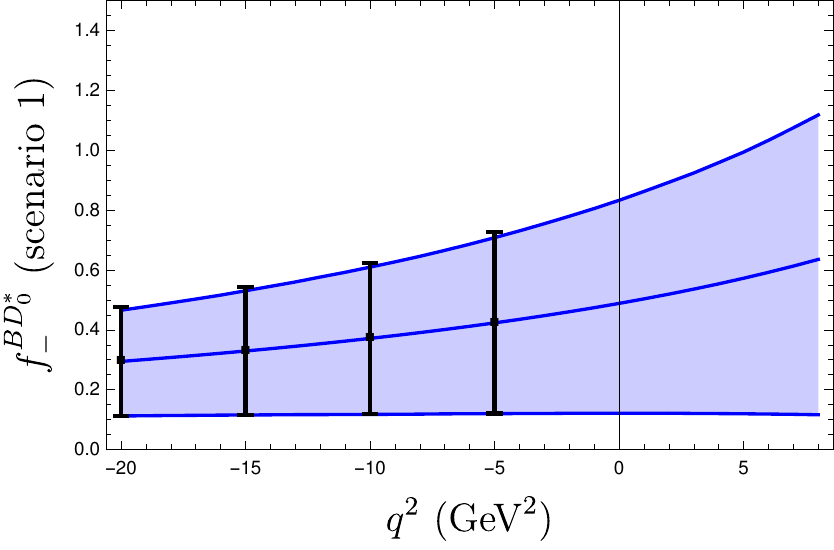}  \\[1.25em]
        \includegraphics[width=\ww\textwidth]{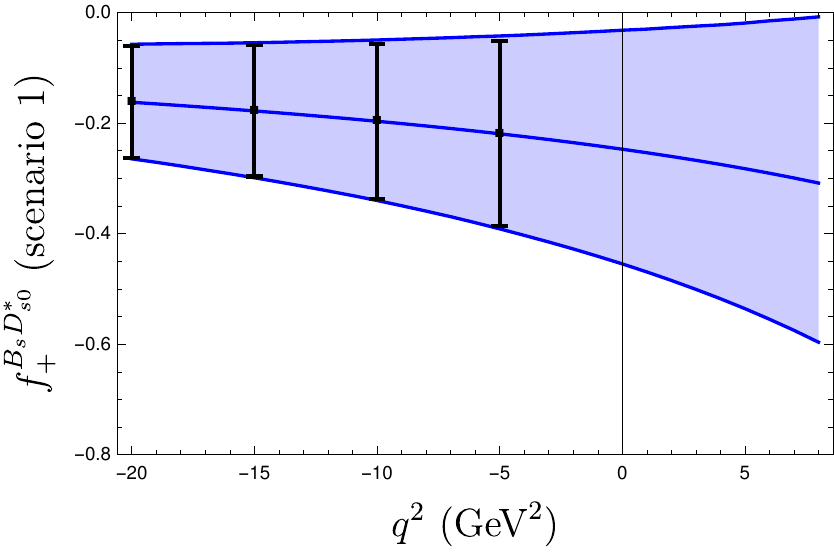}  &
        \includegraphics[width=\ww\textwidth]{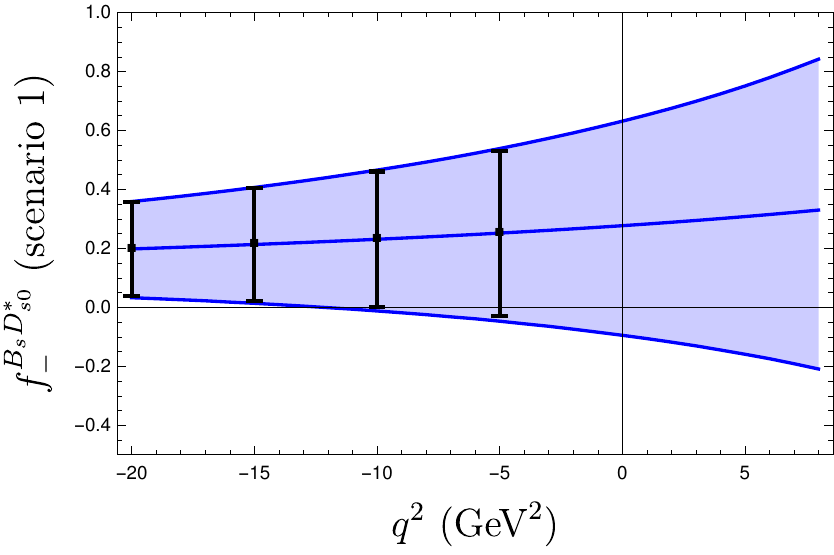}  \\[1.25em]
        \includegraphics[width=\ww\textwidth]{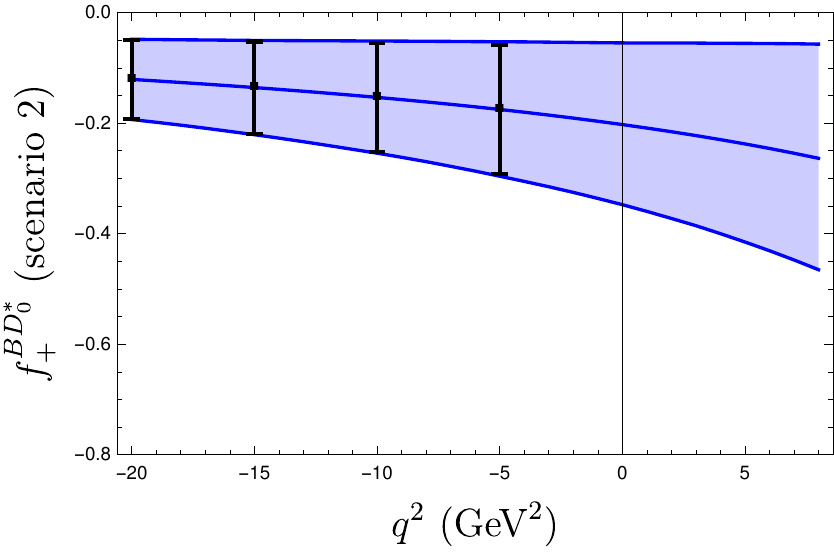}  &
        \includegraphics[width=\ww\textwidth]{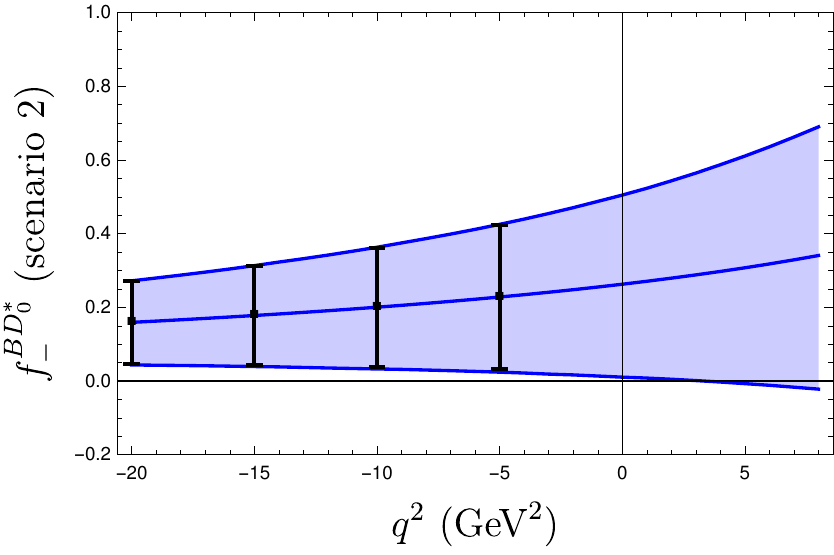}  \\[1.25em]
        \includegraphics[width=\ww\textwidth]{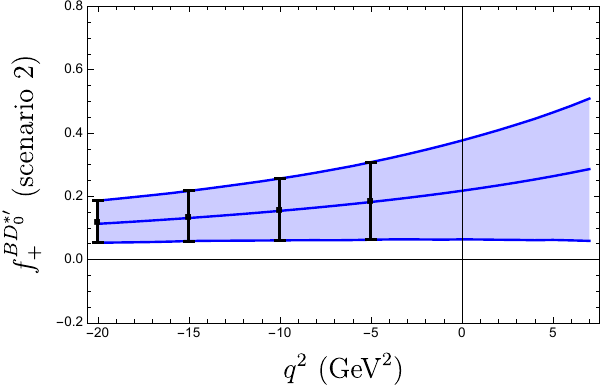}  &
        \includegraphics[width=\ww\textwidth]{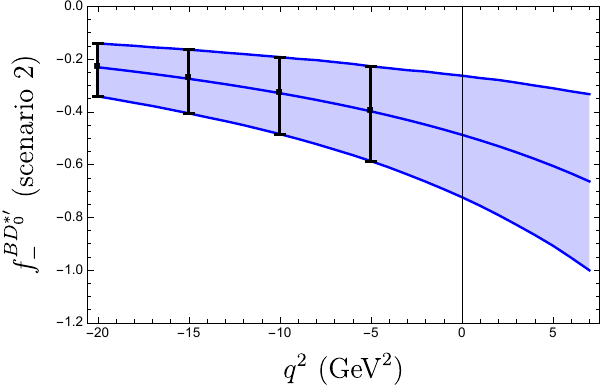}  \\[1.25em]
    \end{tabular}
    \caption{\emph{Form factors plotted as a function of $q^2$. The intervals in black are the results of our LCSR calculation.
    The central values and 68\% probability envelopes in  blue are 
    the results obtained from our fit to the $z$ expansion in \refeq{zexpOPE}. 
    }}
    \label{fig:ffplots1}
\end{figure}

We only consider the first two terms in the parametrization \eqref{eq:zexpOPE}, given the large uncertainties and correlations between the data points used.
Our results for the $\alpha_k^{(i)}$ coefficients are summarised in  \reftab{coeffs}.
With these coefficients, we extrapolate the $z$-expansion (\ref{eq:zexpOPE}) into the semileptonic region,
and finally, via LCSR relations obtain the form factors in that region. 
Our numerical results for the form factors plotted as a function of $q^2$ are shown in \reffig{ffplots1}.
The central values, uncertainties and correlations  of these form factors at any $q^2$ can be easily obtained from  Eqs.~(\ref{eq:LCSRs1fpl})--(\ref{eq:LCSRs2fpmp}), 
using the entries in \reftab{coeffs} and the values of decay constants given in \refsec{SRres}.
In \reftab{ffval}, for convenience of the reader,
we give the values of the form factors at $q^2 = \{0,\,\frac{1}{2} q^2_{\rm max},\, q^2_{\rm max}\}$, where $q^2_{\rm max}=(m_B - m_{D_{0}^{*}})^2$.

The uncertainties of our calculation for  $q^2<0$ further increase in the semileptonic region  due to the extrapolation.
The uncertainties of our LCSR results are mostly parametric and the main contribution to the error budget comes from the $\lambda_B$ parameter.
We stress that a better knowledge of this parameter would significantly improve the precision of the predictions from the LCSRs with $B$-meson DAs, including those obtained here.
\\

\begin{figure}[t]
    \centering
    \newcommand\ww{0.48}
    \begin{tabular}{cc}
        \multicolumn{2}{c}{Scenario 1} \\[0.7em] 
        \includegraphics[width=\ww\textwidth]{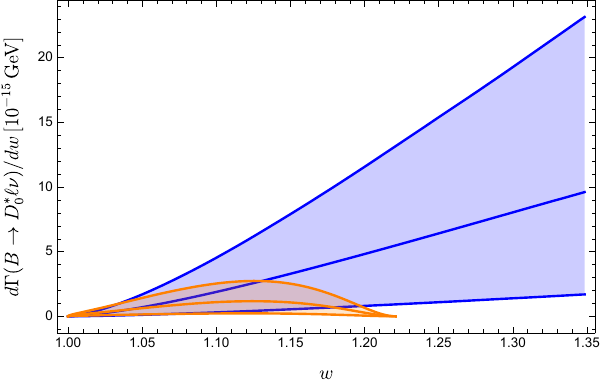}  &
        \includegraphics[width=\ww\textwidth]{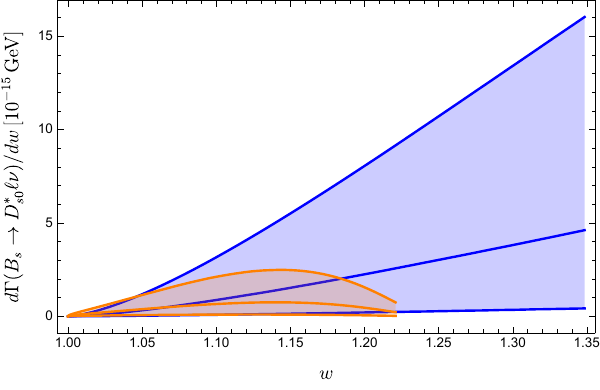}  \\[-0.5em]
        \multicolumn{2}{c}{Scenario 2} \\[0.7em]  
        \includegraphics[width=\ww\textwidth]{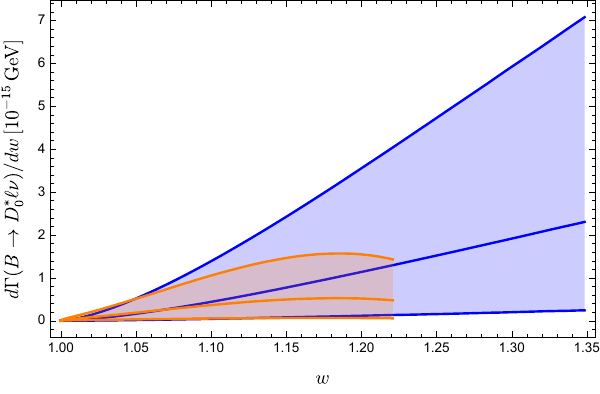}  &
        \includegraphics[width=\ww\textwidth]{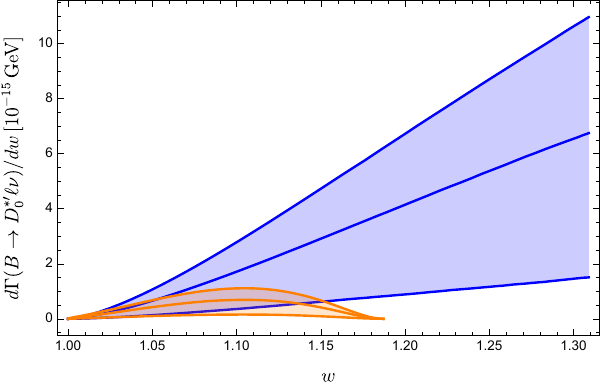}  \\[-0.8em]
    \end{tabular}
    \caption{\emph{Differential decay widths as a function of $w$.
    The blue (orange) bands are the 68\% intervals for $\ell=e,\mu$ ($\ell=\tau$).}}
    \label{fig:widthplots}
\end{figure}

Using our form factor results, we can obtain predictions for physical observables in, e.g., the semileptonic decays $B\to D_0^*\ell\bar{\nu}$, $B_s\to D_{s0}^*\ell\bar{\nu}$, and $B\to D_0^{*\prime}\ell\bar{\nu}$.
We calculate the differential decay widths using the expressions in Ref.~\cite{Bernlochner:2016bci} and plot the results in \reffig{widthplots} as a function of the variable $$ w=\frac{m_B^2-m_{D_0^*}^2-q^2}{2 m_B m_{D_0^*}} \,.$$
For the branching ratios we obtain for scenario 1
\begin{equation}
\begin{aligned}
    {\cal B}( \bar{B}^0\to D_0^* \ell \bar{\nu} ) & = (3.6^{+5.1}_{-3.0})\cdot 10^{-3} \,,\\
    {\cal B}( \bar{B}^0\to D_0^* \tau \bar{\nu} )  & = (3.9^{+5.1}_{-3.1})\cdot 10^{-4} \,,\\
    {\cal B}( \bar{B}_s\to D_{s0}^* \ell \bar{\nu} ) & =  (1.9^{+3.8}_{-1.7})\cdot 10^{-3}\,,\\
    {\cal B}( \bar{B}_s\to D_{s0}^* \tau \bar{\nu} )  & = (2.6^{+4.9}_{-2.2})\cdot 10^{-4} \,,
\end{aligned}
\end{equation}
and for scenario 2
\begin{equation}
\begin{aligned}
    {\cal B}( \bar{B}^0\to D_0^* \ell \bar{\nu} ) & =  (1.6^{+3.2}_{-1.4})\cdot 10^{-3}\,,\\
    {\cal B}( \bar{B}^0\to D_0^* \tau \bar{\nu} )  & = (2.4^{+4.7}_{-2.1})\cdot 10^{-4} \,,\\
    {\cal B}( \bar{B}^0\to D_0^{*\prime} \ell \bar{\nu} ) & = (2.3^{+1.4}_{-1.8})\cdot 10^{-3}  \,,\\
    {\cal B}( \bar{B}^0\to D_0^{*\prime} \tau \bar{\nu} )  & = (1.9^{+1.1}_{-1.4})\cdot 10^{-4}  \,,
\end{aligned}
\end{equation}
with $l=e,\mu$.
We also calculate the lepton flavor universality ratios defined as
\begin{align}
    R(D_0^*) = 
    \frac{\Gamma( B\to D_0^* \tau \bar{\nu} ) }{\Gamma( B\to D_0^* \ell \bar{\nu} )}
    \,, 
\end{align}
for which we obtain in scenario 1
\begin{equation}
\begin{aligned}
    R(D_0^*) &=  0.11_{-0.01}^{+0.03} \,,\\
    R(D_{s0}^*) &= 0.14_{-0.02}^{+0.07}\,,
\end{aligned}
\end{equation}
and in scenario 2
\begin{equation}
\begin{aligned}
    R(D_0^*) &= 0.16_{-0.01}^{+0.04} \,, \\
    R(D_0^{*\prime}) &= 0.077_{-0.010}^{+0.029}\,.
\end{aligned}
\end{equation}
Clearly, in these ratios the theory uncertainties partially cancel and hence their relative uncertainties  are much smaller than in the individual branching ratios.
Our prediction for the branching ratio in scenario 1 agrees with the experimental average 
$\B(\bar{B}^0\to D_0^*(2300) \ell \bar{\nu})= 
(3.0\pm 1.2)\cdot 10^{-3}$
of the Particle Data Group~\cite{ParticleDataGroup:2022pth}.
Also our prediction  for $R(D_0^*)$ (again in scenario 1) is in a agreement with the data driven estimate of Ref.~\cite{Bernlochner:2016bci} (see also Ref.~\cite{Bernlochner:2017jxt}).
We find agreement between our results for the $B_s\to D_{s0}^* \ell \bar{\nu}$ branching ratio and the results of Ref.~\cite{Zuo:2023ksq}, which were obtained with QCD sum rules in the framework of heavy quark effective field theory.
There are no measurements or other theoretical predictions for the observables in the decays of scenario 2.
In fact, our predictions are the first ones for such observables.
With better experimental and theoretical precision, 
the semileptonic decay channels considered here could also be used as an alternative to the well studied $B\to D^{(*)}\ell\nu_\ell$ decays to probe the $b\to c \ell \bar{\nu}$ transitions and to extract $V_{cb}$.

\section{Conclusion}
\label{sec:concl}

We have calculated the $B\to D_0^*$ form factors using for the first time the light-cones sum rules (LCSRs) with the $B$-meson distribution amplitudes (DAs). 
In this method, the $c$-quark mass is finite
and the $B$-meson DAs are defined in HQET. 
This work is complementary to Ref.~\cite{Gubernari:2022hrq}, where similar LCSRs have been applied to the $B$ meson transitions to the axial charmed mesons.
Here, we have also extended the computation  to the $B_s\to D_{s0}^*$ processes taking into account the non-vanishing strange-quark mass.
Our main analytical results are the novel expressions for the LCSRs involving light-cone OPE up to twist-four accuracy for $B$-meson DAs. 
At the same time, we have recalculated the decay constants of charmed scalar mesons from two-point QCD sum rules, which is an important update of earlier works.

The most acute problem related to the charmed scalar  mesons is their identification as resonances in  the $D\pi$ system. 
According to Particle Data Group \cite{ParticleDataGroup:2022pth}, the ground-state  meson with these quantum numbers is $D^*_0(2300)$ which is surprisingly heavy with respect to its strange counterpart $D^*_{s0}(2317)$.
According to Ref.~\cite{Du:2017zvv}, the situation is markedly different and there are two charmed scalar resonances, $D^*_0(2105)$ and  $D^{*\prime}_0(2451)$, where the lightest one  is the natural nonstrange partner of  $D^*_{s0}(2317)$. 
For the sake of completeness, we considered both the possibilities as two different scenarios, and obtained the $B\to D_0^*$ form factors and the $D_0^*$ decay constants for each scenario independently. 

We have calculated the form factor for $q^2<0$ and extrapolated these results to the semileptonic region of the phase space using a $z$ expansion.
Our form factor results have sizeable uncertainties, which are mostly due to the poorly known parameters of the $B$-meson DAs.
We use these results to predict observables for the semileptonic $\bar{B}\to D^*_{0} \ell \bar{\nu}_\ell $ decays.
Our predictions agree with experimental measurements where available.
We also predict observables in the $\bar{B}_s\to D^*_{s0} \ell \bar{\nu}_\ell$ decays.

Furthermore, the results obtained in this paper
allow us to make one more step towards 
filling the observed 
gap between the inclusive $b\to c \ell\nu_{\ell}$ 
decay rate of $B$ meson 
and the sum over  exclusive decay contributions to this rate. 
Having at hand the predictions for 
the partial widths of $B\to D^*_1\ell \nu_\ell$ and $B\to D_0^*\ell \nu_\ell$, 
respectively,  from  Ref.\cite{Gubernari:2022hrq}
and from this paper, we can calculate their 
ratios to the widths of the 
dominant $B\to D^{(*)}\ell \nu_\ell$ modes ,
using the LCSR results obtained for the 
form factors of the latter modes, e.g. from  
Ref.~\cite{Gubernari:2018wyi}.    
These ratios, due to partial cancellation
of DA parameters, will be then more accurate than our predictions for the individual widths. 
Combining the ratios with the well
measured $B\to D^{(*)}\ell \nu_\ell$ widths will 
enable us to estimate the share of 
semileptonic  $B$ decays into charmed axial and scalar mesons in the inclusive rate. 
This type of analysis goes beyond  our scope and will be done elsewhere.

We emphasise that in order to make further 
progress in the $B$ decays involving the scalar charmed mesons, 
it is crucial to have a better knowledge of their spectrum.
Currently, the main information on this spectrum comes from the
studies of the three-body
nonleptonic decay $B \to D\pi\pi $ , identifying  $D_0^*$ resonances in the $D\pi$ two-body subsystem of the final state.
Needless to say, in future the final word on this spectroscopy 
should come from the semileptonic decays $B\to D\pi\ell\nu_\ell$ 
where a careful  analysis of the  observables should help to finally establish 
the mass of the lightest $D_0^*$.
Our results on $B\to D_0^*$ form factors obtained in this paper 
can help to analyse and resolve this problem.

\subsection*{Acknowledgements}

This research was supported by the Deutsche Forschungsgemeinschaft (DFG, German Research Foundation) under grant 396021762 - TRR 257 “Particle Physics Phenomenology after the Higgs Discovery''. R.M. acknowledges the support of this grant via Mercator Fellowship and the hospitality during the visit at Universit\"at Siegen.

\appendix

\section{Light-Cone DAs of $B$-meson}
\label{app:DA}

The definitions and twist expansion of the $B$-meson light-cone distribution amplitudes are taken from Ref.~\cite{Braun:2017liq}. 
These amplitudes are defined in HQET introducing the four-velocity $v=(p+q)/m_B$ of the $B$-meson  with $v=(1,\vec{0}) $ 
in its rest frame. For the two-particle DAs we use
\begin{multline}
\bra{0} \bar{q}^{\alpha}(x) h_{v}^{\beta}(0) \ket{\bar{B}(v)} =
    -\frac{i f_B m_B}{4} \int^\infty_0 d\omega  \bigg\{
        (1 + \slashed{v}) \bigg[
            \phi_+(\omega) -g_+(\omega) \partial_\lambda \partial^\lambda
            \\*
            +\frac12 \left(\overline{\Phi}_{\pm}(\omega)
            -\overline{G}_\pm(\omega) \partial_\lambda \partial^\lambda\right) 
        \gamma^\rho \partial_\rho
        \bigg] \gamma_5
    \bigg\}^{\beta\alpha} e^{-i l \cdot x}
    \Bigg|_{l=\omega v}
    \,.
    \label{eq:2partDAs}
\end{multline}
Here we have defined $\partial_\mu \equiv \partial/\partial l^\mu$ and
\bea
\label{eq:rearr}
\overline{\Phi}_{\pm}(\omega)\equiv \int\limits_0^\omega d\tau
\big(\phi_+(\tau)-\phi_-(\tau)\big)\,,~~ 
\overline{G}_\pm(\omega)\equiv \int\limits_0^\omega d\tau
\big(g_+(\tau)-g_-(\tau)\big)
\,,
\eea
where the DAs of twist-two, three and twist-four, respectively, $\phi_+$, $\phi_-$ and $g_+$, are taken into account. 
Consistently with Ref.~\cite{Gubernari:2022hrq}, we also take into account $g_-$, which is formally a twist-five contribution, in the Wandzura-Wilczek (WW) limit.
Eq.~(\ref{eq:2partDAs}) is equivalent
to the definition in a form of $x^2$ expansion in Ref.~\cite{Braun:2017liq} (see Eq.(4.1) therein), and 
at the same time has some practical advantages, e.g. the barred functions  (\ref{eq:rearr}) are explicitly present.
In our convention, the $B$-meson decay constant $f_B$ in the above definition 
is defined in full QCD and the state  $|B(v)\rangle$ has a relativistic normalisation.

For the three-particle DAs up to twist-four we use:
\begin{eqnarray}
&&
\lefteqn{\langle 0| \bar q(x) G_{\mu\nu}(ux)\Gamma^{\mu\nu} h_v(0) |\bar B(v)\rangle =
\frac{f_B m_B}4 \int_0^\infty d\omega_1\, d\omega_2\ e^{-i (\omega_1 + u \omega_2) v\cdot x}}
\nonumber\\
&&
\mbox{Tr}\biggl\{\gamma_5 \Gamma^{\mu\nu}(1+\slashed{v})
\biggl[
(v_\mu\gamma_\nu-v_\nu\gamma_\mu)  \phi_3
+\frac{i}2\sigma_{\mu\nu} \big[ \phi_3 - \phi_4 \big]
+ \frac{x_\mu v_\nu-x_\nu v_\mu}{2v\cdot x} \big[ \phi_3 + \phi_4 - 2\psi_4 \big]
\nonumber\\
&&
- \frac{x_\mu \gamma_\nu-x_\nu \gamma_\mu}{2v\cdot x}\big[\phi_3 + \widetilde\psi_4 \big]
+ \frac{i\epsilon_{\mu\nu\alpha\beta} x^\alpha v^\beta}{2v\cdot x}  \gamma_5\,
\big[ \phi_3 - \phi_4 +2 \widetilde\psi_4 \big]
- \frac{i\epsilon_{\mu\nu\alpha\beta} x^\alpha}{2v\cdot x}  \gamma^\beta\gamma_5\, \big[ \phi_3 - \phi_4 + \widetilde\psi_4 \big]
\nonumber\\
&&
- \frac{(x_\mu v_\nu-x_\nu v_\mu)\slashed{x}}{2(v\cdot x)^2} \, \big[ \phi_4 - \psi_4 - \widetilde\psi_4 \big]
- \frac{(x_\mu \gamma_\nu-x_\nu \gamma_\mu)\slashed{x}}{4(v\cdot x)^2} \, \big[ \phi_3 - \phi_4 + 2 \widetilde\psi_4 \big]
\biggr]\biggr\}
\ ,
\label{eq:3partDAs}
\end{eqnarray}
where $\Gamma^{\mu\nu}$ is an arbitrary Dirac matrix. In the above,  the functional dependence $\phi_3=\phi_3(\omega_1,\omega_2)$, etc., is not shown for brevity, and the index 3,4 indicates the twist of the DAs.
Also, the Wilson lines are implied but not shown explicitly on l.h.s. of both Eqs.~(\ref{eq:2partDAs}), (\ref{eq:3partDAs}).
This form of the matrix element has been obtained from the one in Ref.~\cite{Braun:2017liq} by defining $x_\mu = z_1 n_\mu$ and $u = z_2/z_1$.

We also introduce the notation 
%
%
for the once and twice integrated three-particle DAs:
\begin{equation}
\overline \Phi (\omega_1,\omega_2) = \int_0^{\omega_1} d\tau\,\Phi (\tau,\omega_2)\ , \qquad
\overline{\overline \Phi} (\omega_1,\omega_2) = \int_0^{\omega_1} d\tau\, \overline\Phi (\tau,\omega_2)\ ,
\label{eq:3int}
\end{equation}
where $\Phi = \{ \phi_3,\phi_4,\psi_4, \widetilde \psi_4 \}$.
The integrated functions defined above enter the OPE expressions,
in particular   the
linear  combinations of the three-particle DAs  multiplying the OPE coefficients in 
\refeq{Iexpan}. These combinations  are defined as:
\begin{eqnarray}
&&\chi_1(\omega_1,\omega_2)\equiv \phi_3(\omega_1,\omega_2)\,, 
\nonumber\\
&&\chi_{2}(\omega_1,\omega_2)\equiv \phi_3(\omega_1,\omega_2)-\phi_4(\omega_1,\omega_2)\,,
\nonumber\\
&&\chi_{3}(\omega_1,\omega_2)
\equiv\overline{\phi}_3(\omega_1,\omega_2)+
\overline{\phi}_4(\omega_1,\omega_2)-
2\overline{\psi}_4 (\omega_1,\omega_2)\,,
\nonumber\\
&&\chi_{4}(\omega_1,\omega_2)
\equiv\overline{\phi}_3(\omega_1,\omega_2)+
\overline{\widetilde{\psi}}_4 (\omega_1,\omega_2),,
\nonumber\\
&&\chi_{5}(\omega_1,\omega_2)
\equiv\overline{\phi}_3(\omega_1,\omega_2)
-\overline{\phi}_4(\omega_1,\omega_2)+
2\overline{\widetilde{\psi}}_4(\omega_1,\omega_2)\,,
\nonumber\\
&&\chi_{6}(\omega_1,\omega_2)
\equiv\overline{\phi}_3(\omega_1,\omega_2)-
\overline{\phi}_4(\omega_1,\omega_2)+
\overline{\widetilde{\psi}}_4(\omega_1,\omega_2)\,,
\nonumber\\
&&\chi_{7}(\omega_1,\omega_2)
\equiv \overline{\overline{\phi}}_4(\omega_1,\omega_2)-
\overline{\overline{\psi}}_4(\omega_1,\omega_2)-
\overline{\overline{\widetilde{\psi}}}_4(\omega_1,\omega_2)\,,
\nonumber\\
&&\chi_{8}(\omega_1,\omega_2)\equiv
\overline{\overline{\phi}}_3(\omega_1,\omega_2)-
\overline{\overline{\phi}}_4(\omega_1,\omega_2)+
2\overline{\overline{\widetilde{\psi}}}_4(\omega_1,\omega_2)\,,
\label{eq:listDA}   
\end{eqnarray}
where the single- and double-barred functions entering
$\chi_{3-8}(\omega_1,\omega_2)$ are defined in 
Eq.~(\ref{eq:3int}).

In the numerical analysis, we use the exponential model (Model-I) proposed in  Ref.~\cite{Braun:2017liq}.\footnote{
    All the DAs are given explicitly in Ref.~\cite{Braun:2017liq}, except for $g_-$, for which we use the model given in Ref.~\cite{Gubernari:2018wyi}.}
It stems from the ansatz  adopted for two-particle DAs in Ref.~\cite{Grozin:1996pq} and extended 
to three-particle DAs in Ref.~ \cite{Khodjamirian:2006st}:
\begin{eqnarray}
\phi_+(\omega) &=& \frac{\omega}{\lambda_B^2} e^{-\omega/\lambda_B}\,,  \\
\phi_-(\omega) &=& \frac1{\lambda_B} e^{-\omega/\lambda_B}
- \frac{\lambda_E^2-\lambda_H^2}{9\lambda_B^3}\, \bigg[ 1 - \frac{2\omega}{\lambda_B} + \frac{\omega^2}{2\lambda_B^2} \bigg]\, e^{-\omega/\lambda_B} \,,\\
%
%
g_+(\omega)&=&
	-\frac{\lambda_E^2}{6\lambda_{B}^2}
	\biggl\{(\omega -2 \lambda_{B})  \mbox{Ei}\left(-\frac{\omega}{\lambda_{B}}\right) +  
	\bigg[(\omega +2\lambda_{B}) \left(\ln \frac{\omega}{\lambda_{B}}+\gamma_E\right) -
   2 \omega \bigg]e^{-\omega/\lambda_{B}}\biggr\}
	\nonumber \\
	\quad &+& \frac{\omega^2}{2{ \lambda_{B}}}\biggl\{1 - \frac{1}{36\lambda_{B}^2}(\lambda_E^2- \lambda_H^2)\biggr\}
 e^{-\omega/\lambda_{B}} ,   \\
g_-^{\rm WW}(\omega)&=&  \frac{3}{4} \omega
 e^{-\omega/\lambda_{B}}  \,\\
\phi_3(\omega_1,\omega_2) &=& \frac{\lambda_E^2-\lambda_H^2}{6 \lambda_B^5}\, \omega_1 \omega_2^2
\,e^{- (\omega_1 + \omega_2) /\lambda_B}\,, \\
%
%
\phi_4(\omega_1,\omega_2) &=&  \frac{\lambda_E^2+\lambda_H^2}{6 \lambda_B^4}\, \omega_2^2 \,e^{- (\omega_1 + \omega_2)/\lambda_B} \\
%
%
%
%
\psi_4(\omega_1,\omega_2) &=& 
\frac{\lambda_E^2}{3 \lambda_B^4} \, 
\omega_1 \omega_2
\,e^{- (\omega_1 + \omega_2)/\lambda_B} \,, \\
\widetilde \psi_4(\omega_1,\omega_2)
&= &\frac{\lambda_H^2}{3 \lambda_B^4}  
\omega_1 \omega_2
\,e^{- (\omega_1 + \omega_2)/\lambda_B} \,,
%
\end{eqnarray}
where Ei is the exponential integral. 
These models depend on three independent parameters
One is the inverse moment $\lambda_B$ of the twist-two DA:
\begin{equation}
\lambda_B^{-1}=\int\limits_0^\infty d\omega \frac{\phi_+(\omega)}{\omega}\,,
\end{equation}
for which we neglect the scale dependence.
The adopted values of this and the remaining two parameters  $\lambda_{E}^2$ and $\lambda_H^2$ are given in Table~\ref{tab:inputsLCSR}
and their origin is explained in Section~\ref{sec:LCSRanalytic}.

\section{OPE coefficients}
\label{app:OPE}

In this appendix we present the expressions obtained 
for the nonvanishing coefficients for the OPE 
calculation of \refeqs{Fexpan}{Iexpan}.

\subsection{Contributions of two-particle DAs}
\noindent
The coefficients for the contributions of two-particle DAs  to $\Fp_{\OPE}$ are:  
\begin{eqnarray}
&&\C_{\phi_+}^{(p,1)}=
\text{mB}-\frac{\text{mc}}{\text{sbar}}\,,
 \nonumber\\
&&\C_{\overline{\Phi}_\pm}^{(p,1)}=-\frac{1}{\text{sbar}}\,,~~~
%
~~~~~~~~~~~~~~~~~~~~~ \C_{\overline{\Phi}_\pm}^{(p,2)}
=-\frac{\text{mc} (\text{sbar}m_B-\text{mc})}{\text{sbar}^2}
\,,
\nonumber\\
&& C_{g_+}^{(p,2)}=\frac{4 \text{mB}}{\text{sbar}}\,,
~~~~~~~~~~~~~~~~~~~~~~\C_{g_+}^{(p,3)}=-\frac{8 \text{mc}^2 (
\text{sbar}\text{mB}-\text{mc})}{\text{sbar}^3}\,,
\nonumber\\
%
&&\C_{\overline{G}_\pm}^{(p,4)}= 
\frac{24 \text{mc}^3 (\text{sbar}\text{mB}-\text{mc})}{\text{sbar}^4}\,,
\label{eq:coeff2p}
\end{eqnarray}
and to $\Fq_{\OPE}$ are:  
\begin{eqnarray}
&&\C_{\phi_+}^{(q,1)}=
-\frac{\sigma\text{mB}+\text{mc}}{\text{sbar}}\,,
\nonumber\\
&&\C_{\overline{\Phi}_\pm}^{(q,1)}=
 -\frac{1}{\text{sbar}}\,,
~~~~~~~~~~~~~~~~~~~~~~~~~\C_{\overline{\Phi}_\pm}^{(q,2)}
=
\frac{\text{mc} (\sigma\text{mB}+\text{mc})}{\text{sbar}^2}
\,,
\nonumber\\
&& \C_{g_+}^{(q,2)}=-\frac{4 \sigma\text{mB}}{\text{sbar}^2}\,,
~~~~~~~~~~~~~~~~~~~\C_{g_+}^{(q,3)}=
\frac{8 \text{mc}^2 ( \sigma\text{mB}+\text{mc})}{\text{sbar}^3}\,,
\nonumber\\
&&\C_{\overline{G}_\pm}^{(q,4)}= 
-\frac{24 \text{mc}^3 ( \sigma\text{mB}+\text{mc})}{\text{sbar}^4}\,,
\label{eq:coeff2q}
 \end{eqnarray}
where $\bar{\sigma}\equiv 1-\sigma$.

\subsection{Contributions of three-particle DAs}

\noindent
The coefficients for the contributions of three-particle DAs  to $\Fp_{\OPE}$ are: 
\begin{eqnarray}
%
&&\C_{\chi_1}^{(p,1)}= \frac{2(1-u)}{\text{sbar} \text{mB} }\,,
~~
\C_{\chi_1}^{(p,2)}=
\frac{ \text{sbar}^2 (1-4 u)\text{mB}^2+3\text{sbar} \text{mB} \text{mc}
    +2 (1-u)(\text{mc}^2-\text{q2})}{\text{sbar}\text{mB}}\,;
\nonumber\\[1ex]
%
%
&&\C_{\chi_2}^{(p,2)}=3(\text{sbar}u \text{mB}- \text{mc})\,;
\nonumber\\
%
&& \C_{\chi_3}^{(p,2)}=\frac{\text{mc}}{\text{sbar} \text{mB}}+
\frac{1}{2}-u\,, 
\nonumber\\
%
&&\C_{\chi_3}^{(p,3)}=\frac{ \text{sbar}(1-2 u)\text{mB}
    \left(\text{sbar}^2 \text{mB}^2 -\text{q2}-\text{mc}^2\right)
    -\left( \text{sbar}^2  \text{mB}^2
    +\text{q2}\right)\text{mc} 
    +\text{mc}^3}{ \text{sbar}  \text{mB} }\,;
\nonumber\\
\nonumber\\
%
%
&&\C_{\chi_4}^{(p,3)}= 6\text{mc}\left( \text{sbar}\text{mB} + (1-2u)  \text{mc}\right)\,;
\nonumber\\[1ex]
%
%
%
&&\C_{\chi_5}^{(p,2)}=
\frac{1}{2}-\frac{\text{mc}}{  \text{sbar} \text{mB}}\,,~~
C_{\chi_5}^{(p,3)}=
\frac{( \text{sbar} \text{mB} -\text{mc}) \left( \text{sbar}^2 \text{mB}^2 +2 
 \text{sbar}\text{mB} \text{mc}+\text{mc}^2-\text{q2}\right)}{ \text{sbar}  \text{mB}
}\,;
\nonumber\\
\nonumber\\
%
%
&&\C_{\chi_6}^{(p,3)}= 6 \text{mc} (\text{mc}-\text{sbar}\text{mB});
\nonumber\\
\nonumber\\
&&
\C_{\chi_7}^{(p,3)}= -\frac{6 \text{mc}^2 (1-2 u)}{ \text{sbar} \text{mB} }\,,
\nonumber\\
%
&&\C_{\chi_7}^{(p,4)}=
\frac{6 \text{mc} \left(- \text{sbar}^3 \text{mB}^3+
(1-2 u)\left( \text{sbar}^2 \text{mB}^2 +\text{q2}\right)\text{mc}   +
   \text{sbar}\text{mB} (\text{mc}^2+\text{q2})-(1-2u)\text{mc}^3 \right)}{\text{sbar}\text{mB}}\,;
\nonumber\\
\nonumber\\
&&\C_{\chi_8}^{(p,3)}=6 \text{mc}\,, ~~
\C_{\chi_8}^{(p,4)}=
18\left(\text{sbar}(1-2 u)\text{mB} +\text{mc}\right)\text{mc}^2\,;
\label{eq:Cfpl3p}
\end{eqnarray}
and to $\Fq_{\OPE}$ are:  
\begin{eqnarray}
&& \C_{\chi_1}^{(q,1)}=
\frac{2(1-u)}{ \text{sbar}\text{mB}} \,,
\nonumber\\
%
&& \C_{\chi_1}^{(q,2)}=
\frac{ \text{sbar} (\text{sbar}(1- 4u)+1+2 u) \text{mB}^2    +3 \text{sbar}  \text{mB}\text{mc} +2(1-u)(\text{mc}^2-\text{q2})}{ \text{sbar} \text{mB}}\,;
\nonumber\\
\nonumber\\
%
&&\C_{\chi_2}^{(q,2)}= -3 (\sigma u\text{mB}+\text{mc})\,;
\nonumber\\
\nonumber\\
%
&& \C_{\chi_3}^{(q,2)}=
\frac{ ( \text{sbar}-2)(1-2u)\text{mB} +2 \text{mc}}{2   \text{sbar} \text{mB}}\,,
\nonumber\\
%
&&\C_{\chi_3}^{(q,3)}=
\frac{1}{\text{sbar}\text{mB} } \Big[ 
-\text{sbar}^2 \text{mB}^3 \sigma (1-2u)- (\text{sbar}-2) \text{sbar}\text{mB}^2 \text{mc}-(1-2u)\text{mB} \nonumber \\
&&\hspace{3cm} \times \left((1+\text{sbar})\text{mc}^2-\sigma\text{q2}\right)+\text{mc}(m_c^2-\text{q2})\Big]\,;
\nonumber\\
%
%
&& \C_{\chi_4}^{(q,3)}=
6 \text{mc} (-\sigma\text{mB} +(1-2u)\text{mc})\,;
\nonumber\\
%
%
&&\C_{\chi_5}^{(q,2)}=
\frac{(\text{sbar}-2)\text{mB} -2 \text{mc}}{2\text{sbar}\text{mB} }\,;
\nonumber\\
%
&&\C_{\chi_5}^{(q,3)}=
 -\frac{(\sigma \text{mB}+\text{mc}) \left(\text{sbar}^2\text{mB}^2
+2\text{sbar}\text{mB} \text{mc}+\text{mc}^2-\text{q2}\right)}{\text{sbar}\text{mB} }\,;
\nonumber\\
%
%
&&\C_{\chi_6}^{(q,3)}=6 \text{mc} (\sigma \text{mB} +\text{mc})\,;
\nonumber\\
%
&&\C_{\chi_7}^{(q,3)}=\frac{6 \text{mc} (\text{mB}-(1-2u)\text{mc})}{\text{sbar}\text{mB} }\,,
\nonumber\\
&&\C_{\chi_7}^{(q,4)}=
 \frac{ 6 \text{mc}}{\text{sbar}\text{mB} } \Big[ \sigma 
 \text{sbar}^2 \text{mB}^3
    +\text{sbar}(\text{sbar}-2)(1-2u)\text{mB}^2 \text{mc}+(1+\text{sbar})\text{mB} \text{mc}^2  \nonumber \\
    && \qquad \qquad \qquad -\sigma \text{mB} \text{q2}
    - (1-2 u)\text{mc}
    \left(\text{mc}^2-\text{q2}\right)\Big]\,;
\nonumber\\
\nonumber\\
&&\C_{\chi_8}^{(q,3)}= 6 \text{mc}\,,~~
\C_{\chi_8}^{(q,4)}=
 18 \text{mc}^2 (-\sigma(1-2u)\text{mB}+\text{mc})\,.
\label{eq:Cfpm3p}
\end{eqnarray}
Here the parameter $u$ is a function of $\sigma,\,\omega_1,$ and $\omega_2$:
\begin{align}
   u = \frac{\sigma\,m_B - \omega_1}{\omega_2} \,,
\end{align}
which induces a  $\omega_1,\omega_2$ dependence in \refeq{Iexpan} for the coefficients at the three-particle DAs.

\end{document}